\documentclass{egpubl}
\usepackage{eurovis2023}

\SpecialIssueSubmission         

\usepackage[T1]{fontenc}
\usepackage{dfadobe}  
\usepackage[compact]{titlesec}
\ifpdf \usepackage[pdftex]{graphicx} \pdfcompresslevel=9
\else \usepackage[dvips]{graphicx} \fi
\usepackage[dvipsnames]{xcolor}
\usepackage[final]{pdfpages}
\titlespacing{\section}{0pt}{*0}{*0}
\titlespacing{\subsection}{0pt}{*0}{*0}
\titlespacing{\subsubsection}{0pt}{*0}{*0}
\usepackage{cite}  
\BibtexOrBiblatex
\electronicVersion
\PrintedOrElectronic
\graphicspath{{pictures/}}

\usepackage{egweblnk}
\title[DASS]%
      {DASS Good: Explainable Data Mining of Spatial Cohort Data} \vspace{-2cm}
\author[A. Wentzel \& al.]
{\parbox{\textwidth}{\centering A. Wentzel$^1$\orcid{0000-0002-2003-2750}
    C. Floricel$^1$,\orcid{0000-0003-0647-9588}
    G. Canahuate$^2$,\orcid{0000-0001-5873-5454}
    M.A. Naser$^3$,\orcid{0000-0003-1020-4966}
    A.S. Mohamed$^3$,\orcid{0000-0002-0777-6202}
    C.D. Fuller$^3$,\orcid{0000-0002-5264-3994}
    L. van Dijk$^3$,\orcid{0000-0002-9515-5616}
    and G.E. Marai$^1$\orcid{0000-0002-7212-9669}
        }  \vspace{-.25cm}
        \\
{\parbox{\textwidth}{\centering $^1$ University of Illinois Chicago, Electronic Visualization Lab
            $^2$University of Iowa
            $^3$University of Texas MD Anderson Cancer Center
       }
} \vspace{-2cm}
} 
\begin{document}

\teaser{
 \centering
 \includegraphics[width=0.9\linewidth]{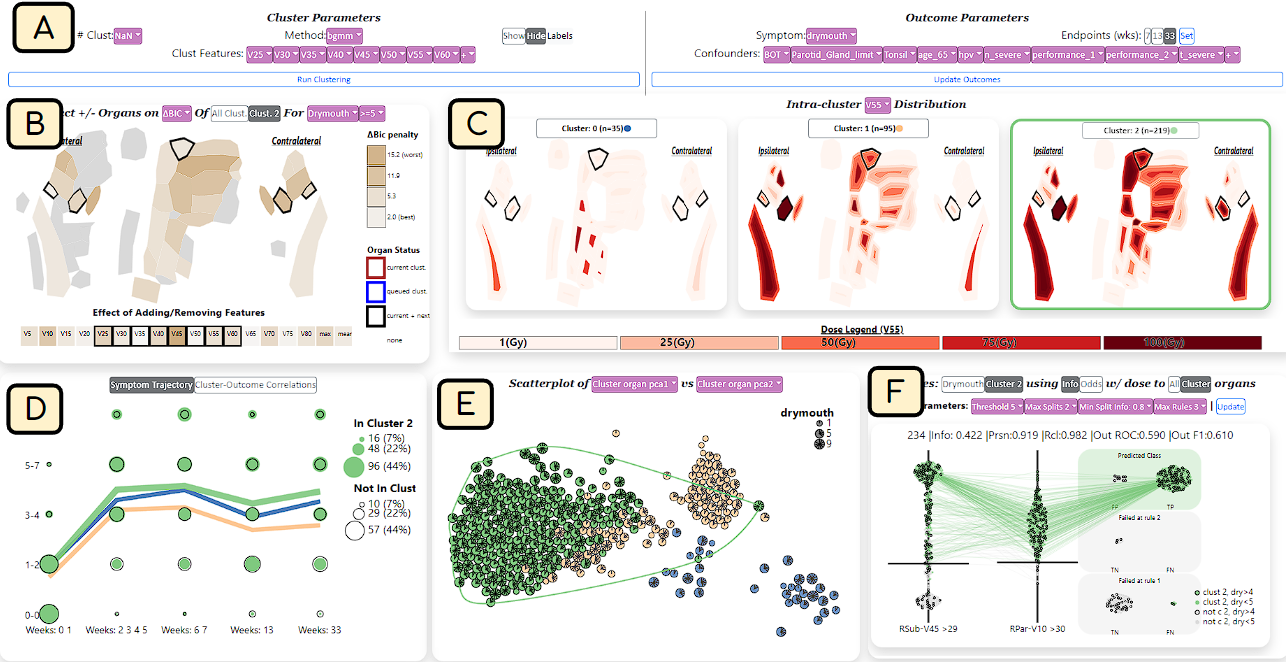}
  \caption{DASS interactive model building for head and neck cancer. A) Control panel for changing cluster parameters and the desired outcome. B) Additive Effect panel showing the effect of changing cluster features. C) Intra-cluster dose distribution plot. D) Outcome plot showing the symptom ratings of patients over time within each cluster. E) Stylized scatterplot showing cohort projections. F) Rule builder view, showing a rule-based mimic model that predicts patients in the selected cluster.}
\label{fig:teaser}
}

\maketitle 
\begin{abstract}  
  Developing applicable clinical machine learning models is a difficult task when the data includes spatial information, for example, radiation dose distributions across adjacent organs at risk. We describe the co-design of a modeling system, DASS, to support the hybrid human-machine development and validation of predictive models for estimating long-term toxicities related to radiotherapy doses in head and neck cancer patients. Developed in collaboration with domain experts in oncology and data mining, DASS incorporates human-in-the-loop visual steering, spatial data, and explainable AI to augment domain knowledge with automatic data mining. We demonstrate DASS with the development of two practical clinical stratification models and report feedback from domain experts. Finally, we describe the design lessons learned from this collaborative experience.

\end{abstract} 

\section{Introduction}
Precision radiotherapy (RT) is a medical paradigm that seeks to personalize cancer RT and care for an individual patient, based on data from cohorts of similar patients. Because for many site-specific cancers, the treatment depends on the location and spread of the disease, modern approaches to precision RT aim to leverage spatial patient-specific information such as anatomical data drawn from CT scans~\cite{wentzel2019cohort,wentzel2020precision}. In conjunction with the cohort data, this information can then be used to improve patient outcomes such as survival or quality of life after treatment. 

In this context, machine learning (ML) models are powerful tools for stratifying the cohort data in meaningful ways, for example into patient groups at high-risk versus low-risk of developing treatment-related symptoms. However, developing applicable clinical ML models for patient stratification is difficult when the data includes spatial information, for example, radiation dose distributions across adjacent organs at risk. In addition, while ML approaches often work well with large oncology data, automated model-building approaches using smaller cohorts often perform poorly when deployed in practice~\cite{marwaha2022crossing}. Furthermore, prediction using treatment plans and qualitative outcomes such as symptom ratings is particularly difficult. This results in simpler models that may underperform or complex models that are very likely to overfit. With advancements in explainable AI techniques, we can better probe models and iteratively find ways of improving models that properly leverage domain knowledge, helping us avoid issues with poor generalization and overfitting, while improving on standard statistical approaches. These combined issues make RT cohort modeling well-suited for a human-machine mixed-initiative system. 

In this work, we present a visual steering approach for creating patient stratifications of head and neck cancer (HNC) patients based on 3-dimensional dose distributions to organs-at-risk, to separate patients at high risk of experiencing long-term side effects. Unlike the current state of the art, our approach supports interactively exploring and visualizing high-dimensional spatial dose distributions, the temporal analysis of RT cohort data, access to both individual patient data and patient distribution within a cluster, constructing unsupervised rule models to help explain the clusters, and iteratively refining and exploring parameters to create actionable stratifications. We implement this approach in \textit{Dose Analytics and Symptom Stratifier} (DASS), a visual computing system designed to allow for the development and exploration of patient stratifications according to different symptoms of interest. We describe two case studies of applying DASS and show how it has been used to improve existing outcome models. Finally, we provide design lessons gained through this collaborative visual steering design.

\section{Background}
Head and neck oncology has seen large increases in patient survival due to a shift from smoking-driven tumors to less aggressive HPV-driven tumors. This increase in survival has resulted in a shift in priorities towards increasing the quality of life of patients: radiation to organs near the primary tumor during treatment can lead to tissue damage, resulting in long-term side effects~\cite{langendijk2008impact, emami1991tolerance}. Predicting when symptoms driven by spatial tissue damage occur is thus an understudied area of interest to oncologists, as it can help identify better treatment guidelines.

When performing the initial diagnosis, oncologists rely on patient history and clinical staging that rank the size and spread of the tumor~\cite{o2016development} to determine the method of treatment to optimize patient survival. However, after the treatment methodology is established, predictive models are needed to identify patients that may need preventative treatment for serious side effects. A diagram of the clinical workflow is available in the supplementary material (Figure A1).

In particular, predicting tissue damage from radiation therapy in head and neck cancer (HNC) patients is a challenge due to the high number of treatment parameters and high number of organs that may factor into side effects. For example, drymouth is often caused by radiation damage to a subset of the salivary glands. Identifying when failure may occur is a difficult modeling task, in which one needs to consider the glands as a spatially interrelated system, as some may compensate for damage to other glands. Additionally, each organ may have a separate non-linear response to the radiation dose over time, and symptom severity varies throughout treatment. Furthermore, the large numbers of HNC patients in a cohort and the dimensionality of the data pose a challenge in terms of visual analysis. Finally, human modelers also require access to individual patient data, as well as to the patient distribution within a cluster to make informed inferences about patient outcomes.

\section{Related Work} 
\subsection{Visual Analysis of Cohort Data} 
Several applications of visual analysis have focused on different algorithmic approaches for clustering patients~\cite{marlin2012unsupervised,mustaqeem2020modular,wentzel2020precision}. Visualization tools often extend these approaches by allowing human-in-the-loop analysis to identify sub-cohorts~\cite{zhang2015iterative,krause2016supporting,bernard2015a}. Other systems have focused on comparison of cohorts to discover differences in disease progression~\cite{malik2015cohort}, genetics~\cite{glueck2017phenostacks}, cancer treatment disparities~\cite{srabanti2022tale}, but, unlike our work, these systems do not focus on model building.

Many systems use clustering~\cite{metsalu2015clustvis} and dimensionality reduction~\cite{oeltze2007interactive,el2008nonlinear} on key features to guide explorations over high-dimensionality data. Some tools have looked at visual analytics for creating clusters with unstructured health data~\cite{kwon2017clustervision,cavallo2018clustrophile,glueck2017phenolines}, while other systems incorporate temporal clustering methods~\cite{zhang2021interactive,zebralla2020obtaining,floricel2021thalis,wang2021threadstates,guo2017eventthread}. However, these systems do not attempt to incorporate spatial information in their clustering models, as we do. Additionally, none of these systems link detailed treatment plans to qualitative patient outcomes in the cohorts, as we do.

\subsection{Visualization of Medical Image Data} 
Work in visual computing with medical imaging often focuses on linking spatial features to external variables to support exploration for domain experts. Early work focused on visualizing spatial imaging data with open source tools (MITK~\cite{wolf2004medical}) and introduced integration of spatial and non-spatial linked views~\cite{gresh2000weave}.

Specialized approaches have been developed to explore cohort features in other domains such as tissue imaging~\cite{falk2018interactive,jessup2022scope2screen,warchol2022visinity}, neuroscience~\cite{jainek2008illustrative,jonsson2020visualneuro,angulo2016multi,ma2018rembrain}, and lumbar spine features~\cite{choi2021dxplorer,klemm2014interactive}.

Focusing on cohorts of RT data, BladderRunner~\cite{raidou2018bladder} visualized cohorts of prostate cancer patients which used a mixture of T-SNE and Gaussian mean-shift clustering to group patients based on bladder shape. VAPOR~\cite{furmanova2020vapor} extended their work to consider RT-induced treatment toxicity. Other work has extended these results to explore uncertainty in RT data for visual analysis~\cite{grossmann2019pelvis,ristovski2014uncertainty} and predictive models~\cite{furmanova2021previs}. However, these approaches do not deal with HNC oncology treatment, which has more complex treatment and symptom patterns but lower temporal variability.

Previous HNC work has used spatial data to cluster patients based on tumor spread to lymph nodes~\cite{luciani2020spatial}. Many techniques rely on simplified representations of anatomical data to allow for better analysis of high-dimensional data~\cite{wentzel2020explainable,wentzel2019cohort,klemm2014interactive,raidou2018bladder}. While these works often deal with feature engineering, none of them focus on directly altering the model in parallel with the visual analysis, as we do. Additionally, we uniquely provide tools for validating the feasibility of the underlying model’s logic and embedding anatomical data directly into the system.

\subsection{Visual Steering and Interactive Machine Learning}
In the medical domain, several projects have developed visualization systems around the workflows of clinical model builders and biostatisticians with a focus on regression models~\cite{dingen2019regressionexplorer}. Raidou et. al~\cite{raidou2015visual,raidou2016visual} proposed a tool for visual analysis of regression-based Tissue Complication Probability models, with a focus on uncertainty. However, these approaches do not focus on clustering or stratification models, as we do.

Other work has focused on actionable explanations for pre-built models for clinicians, such as normal tissue complication models~\cite{zhang2013method}, binary classifiers~\cite{cheng2020dece}, case-based reasoning~\cite{marai2018precision,ming2019protosteer}, and black box models~\cite{cheng2021vbridge}. For explainable AI, DrugExplorer~\cite{wang2022extending} proposed a model for user-centered XAI alongside a system for exploring graph-neural-networks for drug repurposing. However, none of these approaches tackle iterative probing and model development, or capturing spatial information in their data, as we do.

Additionally, our work uses interactive rule mining to help explain the clusters. Many systems have worked on aggregated visualization of rules~\cite{streeb2022task,teoh2003starclass,teoh2003paintingclass,van2011baobabview,muhlbacher2013partition,xu2011rulebender,yuan2021exploration}, and used interactive rule mining to approximate more complex models~\cite{ming2018rulematrix}. Our approach differs from these in that we include a novel rule mining algorithm focused on matching clinical use cases, along with a novel visual encoding that allows for interactive parameter tuning.

\section{Methods}
The DASS design is rooted in our earlier experience with clinical stratification models that relied on forward search for feature selection for clustering~\cite{wentzel2020precision}. Fully automated parameter searches yielded models that performed well on a single performance metric. However, when the clusters were inspected by clinical collaborators, they would often find issues with the organs used, such as organs that are completely unrelated to the outcome, or smaller organs that they felt should be included. Thus, we introduced a human-in-the-loop forward search directly into our front-end alongside model explanations to help improve the process of iterating on our clusters.

User-guided search has two additional benefits. First, our clinician collaborators wished to specify desired characteristics of the models, which led to a need to explore multiple alternative outcomes or starting points based on these desired characteristics. Second, collaborator input is required when balancing model performance, the feasibility of the organs considered, and the number of organs considered. For example, we found that in one model, including both the soft and hard palate had identical effects on the outcome. Thus, the decision came down to the clinicians, who helped us identify which one was of more clinical importance.

Furthermore, in previous work, we attempted to find clusters through hyperparameter search or using predefined cluster features. However, we found that neither approach performed well. Automatic feature selection led to clusters that focus on organs that served as positional \textit{indicator features}, such as the oral cavity~\cite{wentzel2020precision}, but are not causally linked to outcomes and resulted in model explanations that are not well-received. Notably, we found that the brainstem and brachial plexus nerves often appeared as predictors, despite clinicians noting that neither can be associated with any of the outcomes being predicted. Such models work well, but lack causal plausibility which hinders adoption and cannot be generalized to treatment guidelines. The DASS design specifically addresses these problems through its back-end and front-end.

\subsection{Data}
Data were collected from a cohort of 349 HNC patients treated at the MD Anderson Cancer Center using Radiation Therapy, with or without chemotherapy, using a 7-week treatment course. 
We consider three types of data: spatial dosimetric data taken from the patient's treatment plan; unstructured clinical data taken from the patient’s health record; and temporal information on the patient’s self-reported side effects taken during and after treatment. All values are positive ordinal values. Symptom ratings for individuals are discrete, while dose values are continuous.

Diagnostic images were taken at the time of diagnosis, and 40 organs of interest were segmented from these images and considered in the treatment plan. Dose treatment plans were extracted for each organ of each patient. We include 3-dimensional information on the cumulative dose received by each organ during treatment. We use the notation “VX” to denote the maximum dose that penetrates X\% of the organ. For each organ, we consider the V5-V95 range in increments of 5, as well as the mean and maximum dose. 

For outcomes, patients were asked to fill out an MD Anderson Symptom Inventory (MDASI) questionnaire~\cite{rosenthal2007measuring}. This inventory includes self-reported symptoms for 28 different items, such as drymouth and pain on a scale of 1-10. 
We also include secondary variables that may be used as confounders in the patient outcomes taken from electronic health record data, which we generally treat as binary confounding variables.  

\subsection{Collaboration}
Our work was done as part of an ongoing collaboration between data scientists and research oncologists at three US sites. DASS was commissioned to serve first and foremost the needs of the model builders, but to also facilitate clinician input and feedback on the models. Remote meetings were held weekly, during which we would get feedback on designs, and update project goals based on feedback and current results. Examples of prototypes during this phase are included in the supplement Appendix B.

We followed an Activity-Centered Design (ACD) process~\cite{marai2017activity}, which is a methodology conceived to better support designing for domain experts by focusing on existing user workflows and activities. The approach has higher success rates in interdisciplinary settings than Human-Centered Design (63\% versus 25\%)~\cite{marai2017activity}. We focused on the workflows around the development of clinically applicable models, as well as the associated data analysis and verification required to validate and publish the results. 

\subsection{Task Analysis}
\noindent \textbf{Modeling Requirements}
The goal of our project was to aid in the development of an interpretable decision-support tool for clinicians to help identify HNC patients at high risk of long-term severe (self-reported rating > 4 on a 10-point scale) side effects due to radiation damage. We focus on HNC patients as the sensitivity of organs in the head and neck makes detection of quality of life measures in survivors a difficult, under-explored application. Our collaborators were specifically interested in a model that could improve on existing clinical systems by incorporating sets of related organs that together support specific functions, and thus should be treated as a system.

Our system was designed to be used for asymmetric collaborative analysis, which would be handled by model-builders with expertise in the underlying algorithms, with clinicians providing input and feedback. Therefore, we identified requirements for the models themselves, as well as the steps needed to create and validate each model. For our models, we derived the following requirements:

\emph{Actionable:} Usable in a practical setting. In a typical workflow, clinicians use risk stratifications that rank a patient’s risk of survival, which are then integrated into a holistic treatment plan. As such, we require that our models output a simple ranking for each patient, as well as insights that are usable without access to the models. Access to individual patient data, as well as the patient distribution in each cluster, in terms of both doses and symptoms, was necessary.

\emph{Plausible:} Generalize well to a real-world setting. The underlying features that lead to a patient being classified as high-risk must be easy to understand in their spatial context. The models must also place patients in the high-risk group because they received a high dose to a specific set of organs, and the set of organs considered must be mechanistically linked to the the outcome of interest. 

\emph{Transparent:} Be easily probed, assess the plausibility of the models, and identify edge cases in the models. We also needed to be able to demonstrate the plausibility of the models and explain its internal logic to readers with a clinical background.

 
Based on these requirements, we designed a dose-based stratification methodology that clustered 2D dose distributions to a set of organs and used the resulting patient clusters as a proxy for patient risk. Our visual front-end is designed around visual steering, which uses information scent and visual cues to guide our team through the process of selecting, validating, and refining the range of potential parameters for the models to balance different performance metrics and model plausibility. Because this task requires significant knowledge of the models when adjusting parameters, our interactive system is designed to be used directly by models builders and visual computer experts, with encodings designed to allow model builders to communicate intermediate results to clinicians and domain experts.

Through a series of iterative sessions where we developed models and discussed them with our collaborators, we identified the following Activities and Tasks for our visual interface:

\begin{itemize}
    \item \vspace{0em} \textbf{A1} - Given a symptom, find optimal cluster parameters
    \begin{itemize}
        \item \vspace{-.4em} \textbf{T1} Find organs causally related to the symptom of interest.
        \item \vspace{0em} \textbf{T2} Identify a window in the dose-volume histogram that best stratifies the cohort.
        \item \vspace{0em} \textbf{T3} Validate a choice of clustering algorithm and parameters
    \end{itemize}
    \item \vspace{0em} \textbf{A2} - Validate that the logic of a model is causal and plausible
    \begin{itemize} 
        \item \vspace{-.4em} \textbf{T4} Examine the dose distribution of each cluster and where the doses differ.
        \item \vspace{0em} \textbf{T5} Verify if the cluster with the highest symptom risk also has higher doses to the organs used in the clustering.
        \item \vspace{0em} \textbf{T6} Identify confounders that may impact risk prediction.
        \item \vspace{0em} \textbf{T7} Validate the predictive accuracy of the clusters.
    \end{itemize}
    \item \vspace{0em} \textbf{A3} -  Examine and explain individual clusters
    \begin{itemize}
        \item \vspace{-.4em} \textbf{T8} Identify the organ doses that most distinguish each cluster.
        \item \vspace{0em} \textbf{T9} Evaluate differences in symptom trajectory between clusters over time.
    \end{itemize}
\end{itemize}

A1 deals with the development of a models, while activities A2 and A3 help to quantify the models and provide feedback to improve the parameters in A1. A2 is a requirement for clinical publishable findings, while A3 is important for identifying any insights that can be drawn from the final model. For example, once a model is validated, finding that the high-risk cluster for taste dysfunction tends to have a very high maximum dose to the tongue may indicate that future work should investigate the effect of tongue dose on outcomes in more detail.

\subsection{Back-end Algorithms}
\noindent \textbf{Modeling. } \label{modeling}
DASS allows selecting from a range of clustering algorithms: K-nearest-neighbors, Hierarchical clustering, spectral clustering, and a Gaussian Mixture Model. After several iterations, we converged to a Bayesian variant of a Gaussian mixture model for all cluster outcomes. Once a set of organs and a dose-volume histogram (DVH) is identified, these features are encoded as a vector for each patient of size \textit{\#organs * window-size}. Patient vectors are clustered, which are ranked based on the sum of the mean doses to each organ included in the cluster. Ideally, this will result in the highest rank cluster (high dose) being the most correlated with the outcome.

To evaluate the resulting models, we also need to specify a symptom and time point to use as the outcome of interest. We then convert ratings to a binary outcome using a severity threshold. After discussion with our collaborators, the default was a symptom rating above 4 out of 10 at 6 months after treatment. 

Once our clusters and outcomes are identified, we perform multivariate correlation analysis using a likelihood ratio test (LRT) to assess the correlation between each cluster and the outcome of interest, using a set of clinical confounders interactively specified. 

From this, we can calculate an odds-ratio and statistical significance p-value for each cluster, as well as the Bayesian Information Change (BIC)~\cite{konishi2008information}. BIC and AIC are estimates of the goodness of fit of a model that include a penalty for the number of variables considered, in order to prevent overfitting, where lower scores indicate better fits ~\cite{konishi2008information}. For BIC, reductions in score relative to a baseline model of at least 2 indicate reasonable evidence, while reductions of at least 6 indicate "strong" evidence of improvement~\cite{raftery1995bayesian}.  This provides a set of different metrics for assessing the cluster quality in terms of stratifying the cohort.

In addition, to assessing the quality of the current clustering, we provide a forward search in which we alter the existing cluster parameters by adding or removing either a single organ or a single feature from the dose-volume histogram window. We then re-cluster the cohort, and evaluate the new p-value, AIC, and BIC for the new clusters, relative to that of the existing cluster. These metrics are used to provide information scent for users when performing a forward search of the data.

\noindent \textbf{Rule Mining. } \label{rulemining}
To help explain the clusters, we designed a constrained rule mining algorithm and used it to generate a set of dose thresholds that work as a classifier. Our algorithm looks for splits among all dose features in the dataset to find a set that maximizes the mutual information between the splits and a binary outcome. This algorithm is designed to approximate standard rule mining, with the following additional constraints so that the results approximate the rules used by clinicians when specifying dose thresholds: 1) Monotonicity – the high-probability subset for each split in the data must either always be the group above or always in the group below the threshold; 2) Minimalism – The algorithm can only use one dose-feature for each organ; 3) Informative – each “rule” in the ruleset must have a minimal predictive value (user-set) on its own.

Specifically, the algorithm works as follows: 1) we calculate the mutual information gain between each feature split within each ROI (e.g. V40 to the Tongue > 40) and the binary outcome of interest; 2) of the resulting splits, we select the k most important splits; 3) for each of the k best rules, we test combinations of all other splits in step 1 that do not share the same ROI, and calculate the new mutual information gain of the combined rules. Rules are combined using the AND operator (i.e. the patients must satisfy all rules); 4) steps 2-3 are repeated until no improvement is seen in the mutual information gain. To speed up the algorithm, pruning parameters used to speed up the search can be adjusted in the interface.


\noindent \textbf{Implementation}
All data pre-processing and modeling is done using Python with NumPy, Pandas, and Flask for the back-end. Clustering and dimensionality reduction is performed using the scikit-learn package while statistical tests use the statsmodels package. Our system frontend is implemented using React and D3.js.

\section{Front-end Design}
The DASS front-end (Fig.~\ref{fig:teaser}) is composed of 6 panels: a cluster dose view (Fig.~\ref{fig:teaser}-B) that shows the within-organ dose distribution for each cluster (A2), an additive effects view (Fig.~\ref{fig:teaser}-C) that shows the estimated impact of adding or removing features from the cluster on the specified outcome (A1), an outcome view (Fig.~\ref{fig:teaser}-D) that shows the different symptom ratings over time for each cluster (A2-A3), a configurable scatterplot view (Sec.~\ref{fig:teaser}-E) that shows a 2D projection of all the patients in either the dose or outcome space (A2), a rule view (Fig.~\ref{fig:teaser}-F) that shows a set of dose thresholds that best separates a cluster of interest (A1-A3), and a control panel (Fig.~\ref{fig:teaser}-A) that allows users to specify the cluster parameters and outcomes of interest. We arrived at this design following a parallel prototyping process, with multiple design alternatives and repeated feedback. This process is illustrated in the supplemental materials.

To better support the analytical workflow, we use a categorical color scale for cluster membership.  Analysts can select a specific cluster, which is used to populate the temporal outcome and rule views, and brush in all other linked views. By default, DASS automatically selects for brushing the highest dose cluster, as this cluster was typically of the most interest to our clinicians.

\subsection{Visual Scaffolding} \label{scaffolding}

When dealing with organ data, understanding the relative position of each organ is essential for analysis of the relationships between organs and side-effects. Specifically, dose values are correlated with location, and it is important to identify situations where organs may be linked to toxicities due to their centrality and proximity to nearby organs rather than being directly causally linked.

In previous work, we represented the set of organs as a stylized plot showing each organ as a plot in 3 dimensions~\cite{wentzel2019cohort}. However, we felt that this representation was limited in its usefulness, as it is difficult to identify organs that may be smaller and clustered together, but may be functionally important, such as salivary glands and smaller organs in the neck. Previous work has also shown that 2-dimensional maps of anatomical regions work well, and work well with clinicians who are typically trained to work with image slices and 2-dimensional anatomical drawings~\cite{wentzel2020explainable}. Expanding on this, we created a 2-dimensional representation of 45 organs used in our dataset based on existing anatomical drawings~\cite{fehrenbach2015illustrated}.

We then divided up the organs in the head into unilateral organs that sit along the mid-sagittal plane (e.g. tongue), and those that exist as a pair of organs on each side of the mid-sagittal plane (e.g. eyes), which are further subdivided into those on the same side as the primary tumor (ipsilateral side) and those on the opposite side of the primary tumor (contralateral side). This gives us three “groups” of organs along the center axis. For each region, we took tracings around organs of interest using multiple anatomical cross-sections. We then overlaid all drawings, added in missing regions such as the spinal cord, and manually adjusted each contour to avoid overlap and regularize the size of each region. Adjustments were also made to ensure that regions were reasonably concave so that color gradients were visible. A diagram of the final drawing with all regions labeled is available in the supplementary materials.




\subsection{Additive Effects Panel}
\begin{figure}
  \centering
  \includegraphics[width=\linewidth]{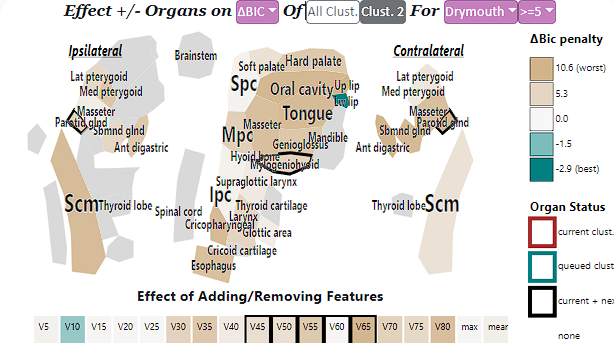}
  \caption{Additive Effects encoding showing a heat map of the organs and dose-features used in clustering. Color encodes the goodness of fit effect of adding (no or teal outline) or removing (dark black or brown outline) features to the clustering. }
  \label{fig:additive}
\end{figure}

When working on model development (A1) our main task is to identify a set of organs to cluster once our desired outcome has been specified (Fig.~\ref{fig:additive}). In this panel, we provide a forward search to estimate the effect of adding (for features not in the current clusters) or removing (for features in the current clusters) different organs or features from the clustering space on model performance (Sec.~\ref{modeling}). We chose a beige-white-teal color scheme as we wanted to de-emphasize uninteresting (negative) results while still capturing the divergent nature of the results. Thus, we used beige as it has lower perceptual salience than the rest of DASS. 

Since model developers may be interested in balancing performance between multiple outcomes, we allow choosing which information metric is used to encode color: BIC, AIC, or the t-statistic–--which we report as a change in p-value, as well as the inputs to the LRT test, and the threshold used to rank an outcome as “severe”.

Alternative designs relied on variations of heat-map and bar charts with effect sizes. However, these were replaced with the visual mapping approach, as we found that it helped to cue users about the approximate position and function of each organ when deciding on clinical relevance. Our collaborators also found that using similar layouts for the dose-cluster encoding and additive effects view reduced cognitive load and made the system more visually consistent.

\subsection{Outcome Plot}
\begin{figure}
  \centering
  \includegraphics[width=\linewidth]{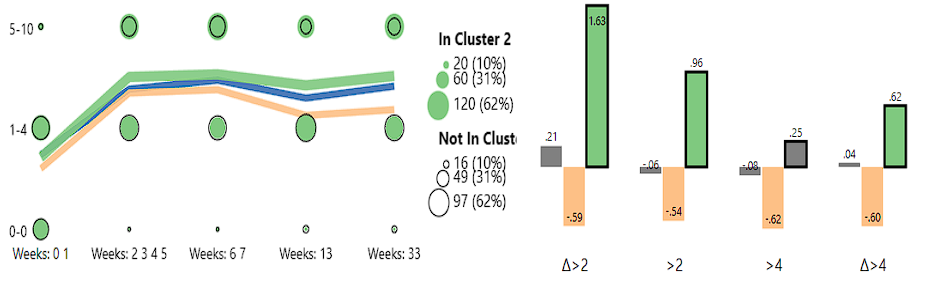}
  \caption{(Left) Plots showing the symptom ratings over time from the start of treatment for the specified symptom of interest, broken up by cluster.  Circular markers encode the percentage of patients that experience a symptom at each level and time  point, and help us estimate a patient's relative risk. Line charts show average ratings for each symptom. (Right) Bar chart showing the results of multivariate correlation tests for the clusters at different thresholds.}
  \label{fig:outcomes}
\end{figure}

To support validation and iterative model improvement, 
it was important to show how outcomes vary within each cluster. This is important when ensuring, for example, that the cluster with the highest doses is actually capturing the high risk patients. To do this, we provide two types of encodings that show patient outcomes for each cluster: a temporal view of symptom ratings for the clusters, and a statistical bar-chart view showing the results of the likelihood ratio tests performed on each cluster for the outcome of interest.

Our temporal view uses a novel encoding (Fig.~\ref{fig:outcomes}) to encode the trajectory of the symptom of interest across the entire treatment period for the patient clusters. This encoding has two components: a symbol grid, and a simple line chart. To reduce the complexity of the encoding, we first group the symptom ratings and treatment dates into bins (we selected five).
In the symbol grid, we divide the patients into those in the selected cluster, and those not in the selected cluster (out of cluster). For each patient, we calculate the highest rating for the symptom within the treatment dates before aggregating by cluster. We then calculate the percentage of patients from the selected cluster that fall in each rating + date bin. These percentages are encoded as circles on a grid, where the x-axis shows each date bin, and the y-axis encodes the symptom ratings. Size encodes the percentage of patients. Values for the in-cluster patients are shown as a saturated marker, while the out-of-cluster patients are shown as a black border marker. By comparing the markers, we can approximate the odds ratio of a patient within the selected cluster having a symptom of a given severity at each time point.

In addition to the symbol grid, we overlay a line chart that shows the mean symptom value over time for each cluster. The line charts use cluster colors. A cluster chart can be clicked to select that cluster for more details.

The statistical bar chart view encodes the results of the LRT test (Sec.~\ref{modeling}). This view is used for assessing how well a model performs while accounting for the specific outcomes and confounders. Cluster-outcome relationships that are statistically significant (p $<$ .05) are shown using their categorical cluster color, while relationships that are not (p $>$ .05) are shown in gray. The selected cluster for the interface is highlighted using a bold black border between the bars for that cluster.

\subsection{Cluster Dose-Distribution Plots} \label{clusters}
\begin{figure}
  \centering
  \includegraphics[width=\linewidth]{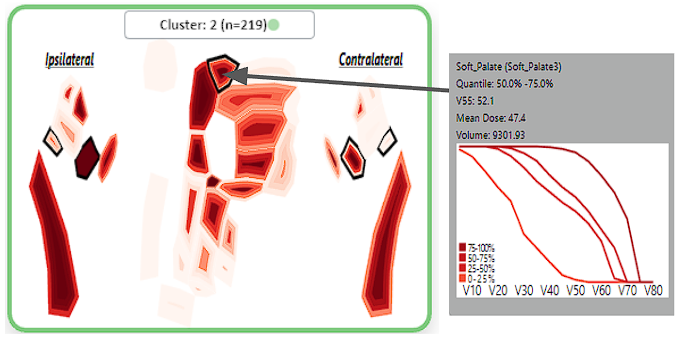}
  \caption{Per-organ dose distribution for a selected cluster. Color gradients shows within-cluster distributions. (Left) A tooltip shows the full dose-volume histogram for a brushed organ. Dotted area shows the value (V55) currently being shown in the heat map.}
  \label{fig:clusters}
\end{figure}

Once a reasonable set of cluster features has been identified, our first set of tasks involves investigating the dose distribution within each cluster (A2 T5-6). This is useful for identifying when the clusters are separating out patients with higher dose to other organs that were not included in the cluster inputs. To do this, we calculate the quartile ranges of a user-selected dose value within each cluster, for each organ. These values are then shown as a gradient heatmap using our 2-dimensional organ diagram using a sequential red color scheme (Sec.~\ref{scaffolding}), where the innermost color represents the top quantile (80\%) and the outer color represents the lower quantile (20\%), allowing us to visualize the inter-organ dose distribution for each organ. 

Interactions allow directly adding or removing organs from the cluster queue, as well as selecting a cluster to be used for brushing in other views. This facilitates the investigation of other aspects of the cluster in more detail. 

To anchor the visual heatmap in the clinician's knowledge, we add a tooltip for each organ that can show the dose-volume histogram for each quantile for the selected organ and cluster(Fig.~\ref{fig:clusters}). This allows for a more detailed view of the entire histogram, while highlighting the relationship between the novel heatmap, and the standard dose-volume histogram that clinicians are familiar with.

\subsection{Scatterplot}
\begin{figure}
  \centering
  \includegraphics[width=\linewidth]{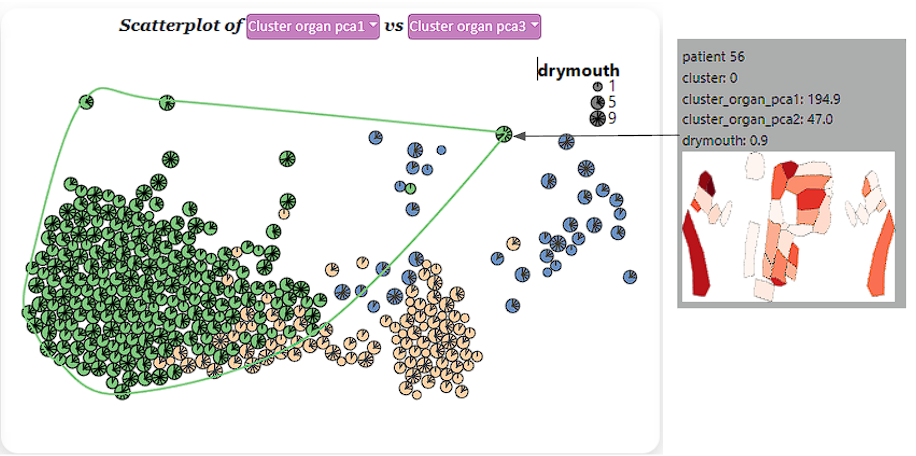}
  \caption{Stylized scatterplot.  Patients are represented by a custom glyph that encodes the outcome of interest (late drymouth ratings) as marks extending radially. Markers are colored by cluster membership, and a contour is shown around the currently selected cluster. A tooltip (left) shows a heatmap of the dose applied.}
  \label{fig:scatterplot}
\end{figure}

To visualize the distribution of patients across each cluster, we include a modified scatterplot panel that shows a 2-dimensional plot of the patients across two interactively-selected dimensions (Fig.~\ref{fig:scatterplot}). 
By default, we show the first two principal components of the features used to cluster the patients, but allow  choosing to alternatively view higher order principal components, the principal components of the symptoms, or individual clinical or symptom ratings. Because we found that avoiding visual occlusion was 
more important than a high-fidelity projection, we use a force directed layout to remove overlap between glyphs.

Each patient in the scatterplot is encoded with a custom glyph that encodes its cluster membership, and the rating for the symptom of interest between 0 and 10. Each circular glyph is encoded with ticks that extend in 32.7-degree intervals in a clockwise radial pattern, where the number of ticks corresponds to the symptom rating. Thus, a full “pinwheel” glyph represents a patient with a symptom rating of 10, while an empty circle represents a patient that does not experience the symptom. Because symptom ratings use discrete ordinal (integer) values, we can encode the exact ratings. We additionally scale the size of the glyph based on the symptom rating to support visual identification of small or high dose values. 

Finally, we color code the glyphs based on their cluster membership. The selected cluster is brushed by giving the corresponding glyphs a higher opacity, and drawing a contour around the convex hull of the cluster in the scatterplot. By hovering the mouse over a patient glyph, the user can view a tooltip showing a plot of the given patient’s received dose, and ratings for all symptoms over time. The dose to each organ is encoded for each patient using the organ diagram heat map  (Sec.~\ref{clusters}).

Previous designs used alternative projection methods with alternative projections and glyph encodings. However, we found that allowing inspection of individuals was more important than preserving location with perfect fidelity. In contrast, T-SNE avoided occlusion, but tended to produce visual clusters that did not correspond to the desired clusters. For glyph design, we considered alternative shapes (e.g. diamonds or circles) for different levels of severity. However, collaborators found the use of color and shape confusing, while the use of ticks + size was better received and we were able to identify the patient of most interest (very high and very low severity) fairly easily for further inspection.

\subsection{Rule Builder}
\begin{figure}
  \centering
  \includegraphics[width=\linewidth]{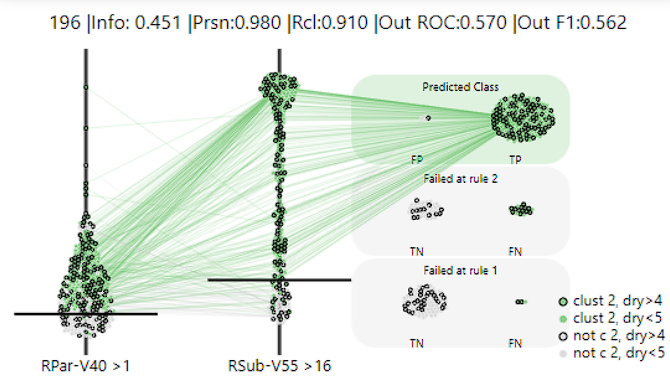}
  \caption{Ruleset encoding in the rule mining view. A swarm plot of the patients is shown for the feature used in each rule, with the first and most informative rule on the right. A horizontal line shows the cutoff thresholds used in the rule. Patients that pass a rule are then plotted in a swarm plot in the next rule on the right. The section on the right shows rule patients failed at, with patients that pass all rules at the top (green area). Patients in each section are divided to show the False Positives or False Negatives at each level. Lines connect markers for a patient across each sub-plot.}
  \label{fig:rules}
\end{figure}

Once our clusters are built, one of our goals is to explain the clusters in terms that are familiar to clinicians. To accomplish this, we used a constrained rule mining algorithm (Sec.~\ref{rulemining}) to produce a set of dose thresholds such that the group of patients that meet these thresholds approximates the selected cluster. This approach was chosen as clinicians often work with dose thresholds when choosing treatment plans. 

When a cluster of interest is selected, our algorithm finds a list of rule-sets that optimize the mutual information between the patients and the cluster of interest. We then generate a plot for each ruleset, and show the top rules in a list to the user. We also show the number of predicted positives, information gain, precision, recall, and f1 for predicting the true class above each plot. 

Our novel rule encoding is based on a mixture of swarm plots and parallel coordinate plots that are modified to show the progressive filtering of each ruleset (Fig.~\ref{fig:rules}). We encode each feature (e.g. V50 to the tongue) along the x-axis. We then map the y-axis to the dose value in grays. Patients are  plotted along the y-axis based on their value for the given dose feature in the x-axis, and adjusted using a force-directed layout to avoid overlap. A horizontal line is then drawn at the threshold of the rule for the feature on each step of the x-axis. Patient marks are color-coded based on the selected cluster, while patients not in the selected cluster are gray. 

To show the effect of additional rules, the features along the x-axis are ordered from left to right by the maximum information gain for its corresponding rule. In the first feature, we show all patients in the cohort. For additional features, we filter out all patients that do not satisfy rules from all previous features. The rightmost side of the encoding shows the patient groups stratified along the y-axis based on when they were filtered out of the ruleset. The set of patients that satisfy all rules is grouped at the top, while the set of patients that do not satisfy the first rule is grouped at the bottom. We further separate the final group by those in the true class (target cluster) and those not in the true class, allowing us to visualize the false positives and false negatives for each rule.

To provide a visual cue for how the rules are filtering the cohort along the x-axis, we provide lines that connect the undistorted locations of patients between axes, equivalent to a parallel coordinate plot with filtering. Once a patient is filtered out, we draw a line from the corresponding rule to the group on the right side. To prevent overlap, we only show the lines for the patients within one stratum at time, which is changed by brushing a patient in the given strata. By default, we brush the group of patients that satisfy all rules (predicted positives).

\section{Evaluation}
The first and foremost value of DASS comes from its unique functionality and its ability to support clinical model development, which we illustrate via two case studies. These case studies, presented here in abbreviated form, illustrate the process of creating models for practical use, based on real clinical data. The case studies were performed via Zoom meetings with desktop sharing, with one of the data scientists piloting DASS and the group using the think-aloud methodology with note-taking. We furthermore collected and report qualitative feedback from clinical collaborators during these case studies.

As further evidence of the DASS functional value, we provide in the supplemental materials a quantitative evaluation of clusters generated with DASS against baseline ML clusters. The DASS clusters improve performance for drymouth, choking, and swallowing issues. Finally, with an eye towards the generalizability of DASS to other modeling problems, we collected additional feedback where eight data scientists 
rated the usefulness and usability of DASS.

Since the interactive model-building components are directly targeted at modelers,  An additional quantitative comparison of our clusters against baseline ML clusters generated without DASS can be found in the supplemental materials.

Our dataset consists of 349 patients treated with radiation therapy for oropharyngeal cancer. These models have been generated with the help of DASS by four data scientists in our group over several months of remote collaboration. The models have shown improvements over baseline models, and have been favorably evaluated by three clinical oncologists.

\subsection{Case Study 1}
\begin{figure}[htb]
  \centering
  \includegraphics[width=\linewidth]{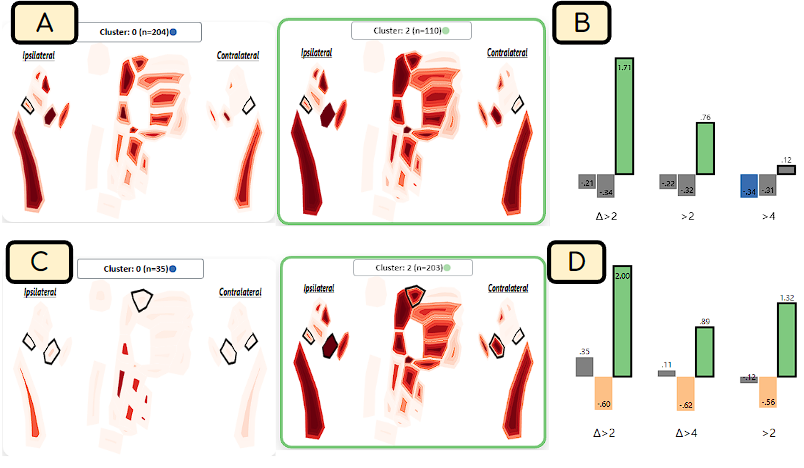}
  \caption{Case 1. (A) Low-and high-dose clusters using starting features.  The low dose cluster includes several organs with high variance in the dose distribution. (B) Initial model performance. (C) Low- and high-dose clusters using the final model. Low dose cluster has a much lower variance with only a few sets of outliers. (D) Final model performance measures. High-risk cluster is correlated with drymouth with a higher odds ratio than the initial clusters.}
  \label{fig:case1}
\end{figure}
Our group was interested in identifying patients at high risk of developing drymouth at 6 months after treatment, a common side effect in HNC patients. In particular, the clinician analysts in the group wished to model the relationship between drymouth and the radiation dose applied to the salivary glands. The medical literature had established a few dose guidelines for parotid glands, but not for other salivary glands. 

The model building process started by setting the parameters in the DASS control panel. Based on results from earlier work~\cite{wentzel2019cohort}, the group set the initial clustering features to be V40-V55 doses to the ipsilateral and contralateral Parotid glands. Three clusters based on a Gaussian mixture model were investigated. Inspecting the initial clusters in the outcome plot, the analysts noticed that, as expected, there was a higher rate of drymouth in the highest dose cluster (Cluster 2 in Fig.Sec.~\ref{fig:case1}), although the correlation was not significant for the desired threshold of $>$ 5. Moving to the dose distribution plot, the group noted that the low and medium dose clusters tended to have a high-variation in the dose to certain organs, as indicated by the dark red inner contours and light outer contours to several organs (Fig.~\ref{fig:case1}-A), suggesting that the model parameters did not differentiate the low dose patients well. Moving to the additive effects view, the model was iteratively adjusted to include the submandibular glands and soft palate, with a larger dose window (V30-V55). After updating the model, the group noticed the clusters in the scatterplot panel achieved much better separation in the data (Fig.~\ref{fig:case1}-D) compared to using just the parotid glands (Fig.~\ref{fig:case1}-B). Returning to the dose cluster plots, the group also verified that the low dose cluster had a lower overall variance in the doses (Fig.~\ref{fig:case1}-C).

Once the group achieved a set of features, the analysts aimed to verify the validity of the resulting model. Looking at the outcome panel, they noticed that while the high dose cluster was a strong predictor of drymouth, the low dose cluster had a high odds ratio. Moving back to the scatterplot, and with the help of the oncologists, they inspected the patients in this low dose group, and noticed an interesting pattern: a number of patients had very high symptom ratings, and confirmed that none of their organs received notably high doses. Pivoting to the temporal outcome panel, the analysts further noted that this low-dose group had the highest incidence of severe drymouth at the start of treatment. After further discussion with the clinical collaborators, the group concluded that existing treatment plans try to minimize dose to the parotid glands, but not the submandibular glands, so the dose tends to be much lower in severe cases. The team theorized that there is likely a minor, but not full compensatory effect of the contralateral salivary glands when one set of salivary glands fails that should be explored later when investigating dose guidelines.


\subsection{Case Study 2}
\begin{figure}[htb]
  \centering
  \includegraphics[width=\linewidth]{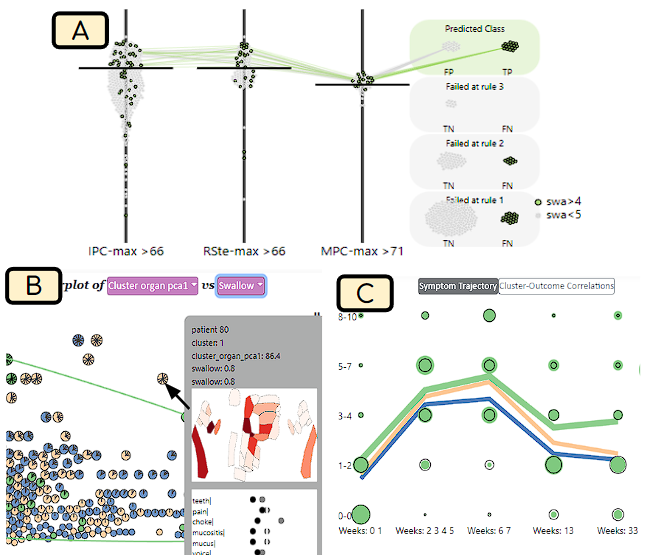}
  \caption{Case 2. (A) Rule mining results for predicting severe swallow dysfunction, which suggest using high doses to the pharyngeal constrictors.  (B) Scatterplot of the first principle component of the cluster features vs swallow ratings. A tooltip highlights a case with severe swallowing in a low-dose cluster.  (C) Outcome plot for the final clusters. High risk patients have similar ratings during treatment, but swallowing issues increase between 6 weeks and 6 months after treatment.}
  \label{fig:case3}
\end{figure}
This second case study dealt with the identification of patients at high risk of swallowing dysfunction, which is a less common outcome that is theorized to be related to damage to muscles in the mouth and throat. Swallowing disorders are also related to patients that require a feeding tube and weight loss, and thus it is an important outcome to avoid. High-risk patients can also be assigned prehabilitative therapy such as swallowing exercises as well.

To help identify a set of starting organs, the analysts inspected the rule mining view and set the desired outcome to be severe late swallowing using all available features (Fig.~\ref{fig:case3}-A). By looking at the resulting rules, the group was able to identify the organs and dose features that best predicted severe swallowing, which allowed selecting a set of starting features for the cluster. Among the best splits were high dose depths (V55-V70) to the superior, medial, and inferior pharyngeal constrictors, which are key muscles used in swallowing, which were chosen as a starting point for the clusters. After running the clustering, the analysts inspected the outcome view and noticed that the initial clustering parameters were effectively separating the high-risk patients: this highest dose cluster had a significantly higher odds ratio of severe late swallowing (2.56) than other clusters (Fig.~\ref{fig:case3}-C). Inspecting the cluster dose distribution view, it was noted that this high-dose cluster was noticeably smaller (n = 35) than the drymouth cluster and that the high-dose cluster tended to consistently have a much higher V55 to the IPC than other clusters.

Moving to the scatterplot, the analysts changed the dimensions to show the first principle component of the dose and swallowing ratings, which allowed identifying all patients with high swallow dysfunction that were not in the cluster (Fig.~\ref{fig:case3}-B). Using the tooltip, the group found some of these patients had high doses to the base of the tongue and upper larynx. The analysts then added the supraglottic larynx to the clustering parameters in hopes of capturing this group. The group then moved to the additive effects view, iteratively changed the dose window to include only the V55-V65, and added the esophagus, which is another major muscle used for swallowing in the base of the throat. After finalizing the parameter set, the analysts inspected the rule view to find the features that best distinguished the high-risk cluster. This high-risk cluster was easily distinguished using the V55 to the Inferior Pharyngeal Constrictor. Our clinical collaborators noted that all the pharyngeal constrictor muscles are located close together, and there exist guidelines for the dose for all of these muscles. Thus, a high IPC dose is likely a predictor of a high dose to all related organs. Additionally, the group discussed the fact that the dose threshold for swallowing was higher than drymouth, which may indicate that muscles are less sensitive to radiation relative to salivary glands.

\subsection{General Usefulness and Usability Feedback}

\begin{figure}[htb]
  \centering
  \includegraphics[width=\linewidth]{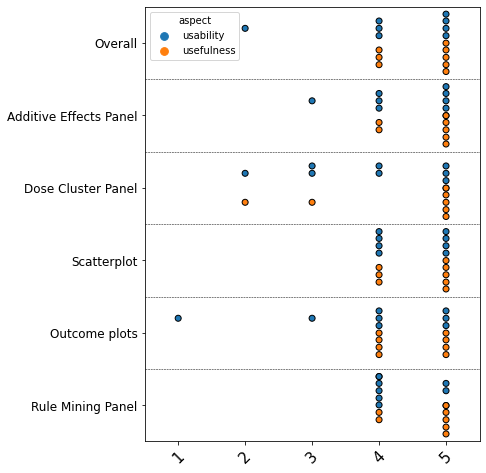}
  \caption{General DASS usability and usefulness.}
  \label{fig:usability}
\end{figure}

In addition to the case studies, which illustrate the DASS unique functionality, we collected qualitative and quantitative feedback from both collaborators and from modelers not affiliated with the project. All collaborators appreciated the functionality provided by DASS, and are in the process of publishing the resulting clinical models. Regarding the spatial cluster panel, our clinical collaborators found it intuitive and useful for inspecting dose distributions of organs of interest. Feedback on the rule mining algorithm was also positive, with oncologists remarking that it was “very useful”, as it could “translate our results into practical applications”. A data mining expert responded similarly to the additive effects panel, saying that it was a “nice, very nice way to explore the parameter space”.

Additionally, we asked, via an anonymous online questionnaire, three senior data scientists in the group, who were not directly involved in the DASS design but participated in walkthroughs of the system, and five junior data scientists, who were not affiliated with the project to rate the usefulness and usability of the whole system and of each component of the system on a Likert scale from 1 to 5. We specifically sought feedback from data scientists, with an eye towards generalizability, as modelers are the intended users of the interactive model-building components of the system. Results are shown in Fig.~\ref{fig:usability}.

Feedback was very positive, with most ratings between 4-5, in particular for usefulness. Ratings for usability were slightly lower, as expected: some of the group experts clarified the group’s narration during the model-building process was extremely useful, and they wished for visual help buttons replicating that experience on demand. Ratings from the junior data scientists not affiliated with the project were occasionally lower, in particular for the composite outcome marker plots and the dose cluster panel. Based on the qualitative feedback, the difference in these cases was directly related to the visual scaffolding and domain expertise which collaborators benefited from, as these plots were based on methods used in RT planning and clinical biostatistics.

\section{Discussion and Conclusion}
Our design relies on three main principles for improving model development: 1) information scent to guide model development (A1); 2) visual scaffolding to support bridging the information gap between what domain experts commonly deal with and what is needed to reason about the data (A2); 3) model explanations aimed at translating our novel approach to the types of simpler “models" use in practice (A3). Our case studies show how the system was effectively used to develop explainable models that outperformed our previous attempts at developing clinical models.


Below, we distill the design lessons gathered from this project when dealing with visual steering and explainable AI problems in collaboration with domain experts.

\noindent \textbf{L1.} \textit{Explanation Scaffolding:} We extend the concept of visual scaffolding – gradually building to more complex visualizations from a more familiar one – to that of XAI-style model explanations. Specifically, we argue that model explanations should aim to translate more complex models into those that mimic how users commonly deal with the data. In our case, we used constrained rule mining in conjunction with visualizing intra-cluster dose distributions using a visual scaffolding approach. Other systems have used regression models which are common in biostatistics. However clinicians do not often reason about such models directly, so they are less useful in clinical practice.

\noindent \textbf{L2.} \textit{Keep Model Goals Flexible:} When developing models, data scientists may work solely to optimize the performance in terms of easily measured outcomes~\cite{mao2019data}, which leads to issues during collaboration with model end-users~\cite{zhang2020data}. In practice, there is often a misalignment between what can easily be measured, and what makes a model useful in practice. In developing our models, we found that it was important to allow users to investigate a mixture of outcomes, in addition to qualitative factors such as model plausibility and complexity, which need to be leveraged against each other when deciding on the final model.

\noindent \textbf{L3.} \textit{Encourage Skepticism:} One motivation in the design of our system was a recurring problem of designing models that performed well, whereas further probing revealed internal logic that appeared to be the result of biases and spurious correlations in the data. Despite this, our models were often received without skepticism when these issues were not brought up. This issue with over-trusting erroneous explanations has been suggested in early empirical studies~\cite{kaur2020interpreting,xiong2019illusion}. The communication gap between model builders and experts may result in dramatically over-trusting the models for both parties as they may be unable to identify issues in the models on their own. When dealing with XAI, designers should focus on promoting skepticism about the models by highlighting potential issues in the models, such as outliers and confounders, which can help highlight previously unknown issues in the models.

The main limitation of our system is the reliance on visualizations that require familiarity as well as knowledge of the underlying models and data, which is made possible by the long-term nature of our collaboration. While we use domain-specific designs for our visual scaffolding approach and model designs, the design philosophy can be generalized to other problems involving spatial data where model outputs can be translated into discrete groups, such as clustering and decision trees. 
In terms of scalability, our system requires 5-15 seconds to update new results for each cluster, depending on the number of clusters and rules mining settings. Scaling to larger datasets may increase the required time, although this is still significantly faster than alternatives that do not use interactive steering. Visualization of individual patients in the Scatterplot and Rule view may also be difficult with very large cohorts.


In conclusion, we have presented an ML and visual steering system for clinical oncology symptom modeling with spatial data. We described the co-design of a clinical visual-steering system, and demonstrated its ability to support the creation of interpretable ML models for stratifying patients. Additionally, we presented a set of lessons learned for model co-development and model explanations for a hybrid, machine expert and human expert problem. We hope that these findings will help future designers create better, and more trustworthy models in high-stakes settings.

\noindent \textbf{Acknowledgements}
\noindent Our work is supported by NIH awards NCI-R01-CA258827 and NLM-R01-LM012527, and NSF awards CDSE-1854815 and CNS-1828265.

\bibliographystyle{eg-alpha-doi}
\bibliography{egbibsample}

\newpage
\section{}

\setlength{\voffset}{0cm}
\setlength{\hoffset}{0cm}



\section*{Supplementary material}

# A. Workflow and Background

Fig A1. Physician Workflow. The main items of interest for this project are to help establish a stratification of patients' risk of certain side-effects from their radiation plan (Side-Effect red box), which can then be used to identify which patients require additional preventative treatment, as well as help identify dose thresholds (Dose thresholds red box) that can be input as soft-constraints during treatment planning.

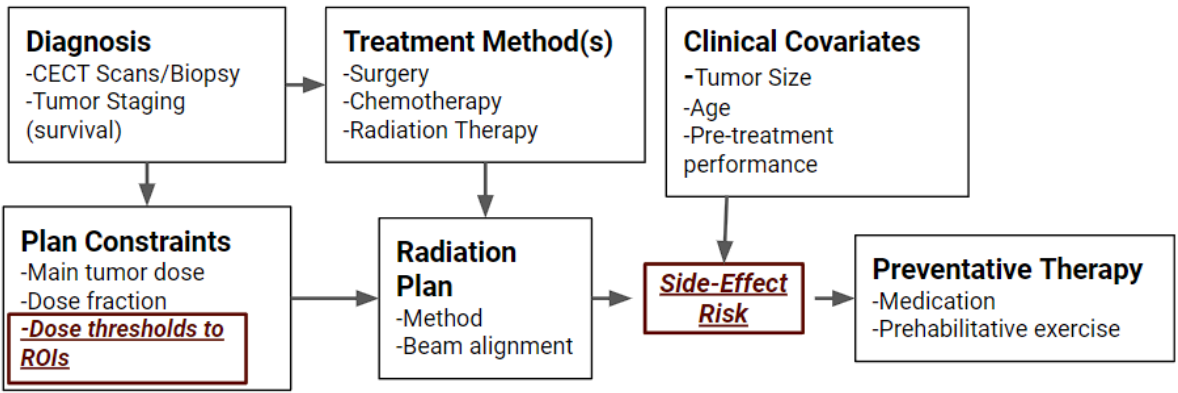

Fig A2. Full 2D Mapping of Regions of Interest. Regions represent approximate locations of important anatomical structures along the center, and sides of the head. The relative size of smaller regions are exaggerated to improve visibility.

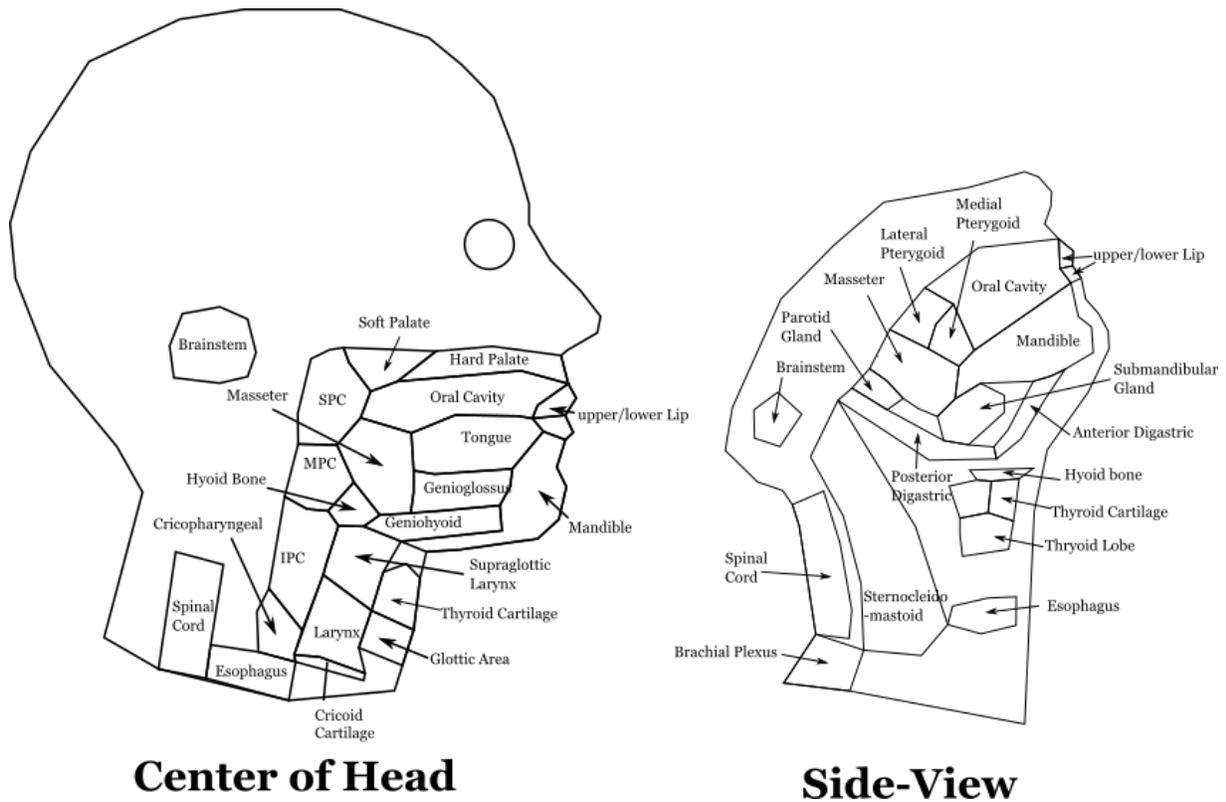

Fig A3. DASS Workflow. First, a desired outcome is selected, along with the initial clustering parameters, which can be drawn from prior literature or informed from information metrics in the explanation view. After clusters are generated, the inter-cluster distributions can be investigated using the dose distribution and configurable scatterplot views. Cluster performance can then be validated by investigating the inter-cluster symptom trajectories and correlations with outcomes. A rule-based classifier can also be used to produce explanations for the high or low-risk clusters based on dose thresholds. Once clusters are investigated, the Additive Effects panel can identify potential changes to the clustering parameters that could improve the model performance.

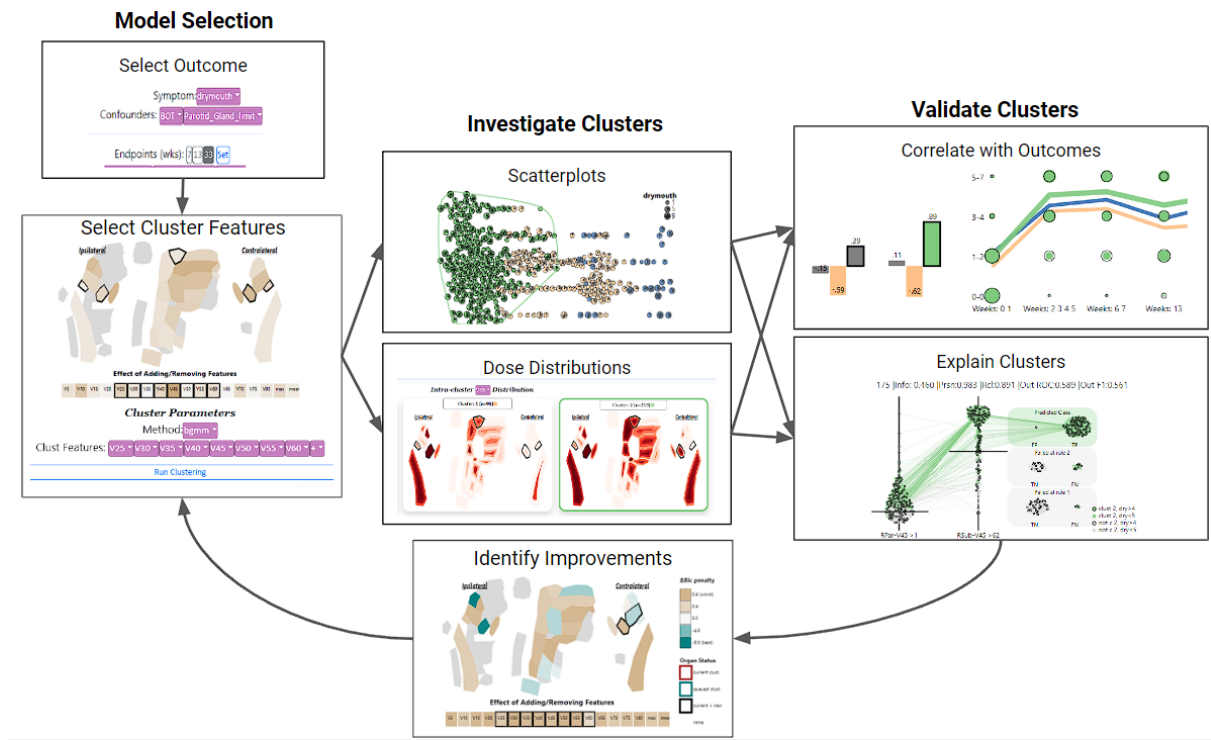

## B. Prototypes and Designs

This section details some of the early prototypes and approaches that led up to our final design of DASS, which covers both algorithmic and visualization attempts.

Fig B1. Early prototype of the visual scaffolding interface used to inspect the results of clusters generated using grid-search or physician-selected features without visual steering. Each row shows the distribution of mean-dose to each ROI within each cluster. Bar charts on the left show odds-ratios of different symptoms. ROIs used in the cluster use a saturated red color scheme, while ROIs not used in the clusters use a desaturated grayscale color scheme. This was our original approach that inspired the design of MOTIV, as we felt that our clusters were not producing outcomes that achieved both good performance and explanations that aligned with clinical knowledge.

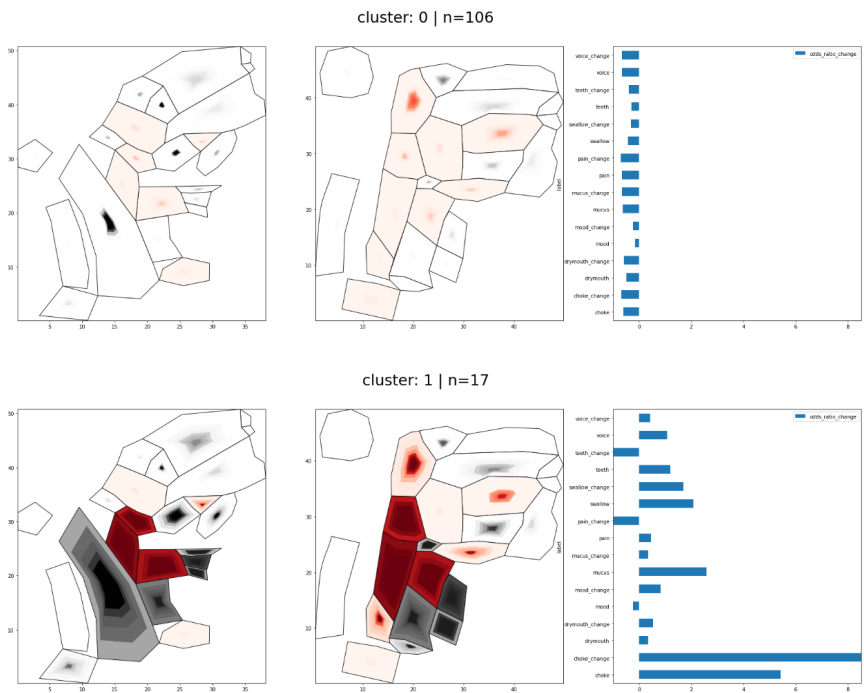

Fig B2. Alternative Approach to producing cluster explanations. For each cluster, we trained a spline-regression model to predict cluster membership using the dose-values used in the cluster. For each roi we selected the dose-value (e.g. V15, V20) to the roi that was the most strongly correlated with the cluster using a Chi-square test. Each row represents a cluster. Charts on the left are the shape-curves for each input to the spline model. Bar charts on the left encode the parameter coefficients for each dose-value.

This approach was abandoned as the model results do not correspond with the expected behavior, in which all shape curves should be monotonically decreasing or monotonically increasing.

Fig B3. Graph of an earlier attempt at creating a clinical risk model. Dose clusters were generated using pre-defined groups of organs based on region or function. A constrained causal learning model (Fast Greedy Equivalent Search) was then run on the cluster memberships, clinical covariates, and severe symptom ratings at different time points, as well as membership in clusters of symptoms taken from prior work. This approach was implemented in hopes of establishing causal clusters, but was abandoned as it was difficult to consistently generate results that were useful for clinical workflows.

Fig B4. View that shows the effect of each cluster on a machine learning model for predicting different presence of a symptom at 6 months at different thresholds. A user selected model is trained to predict each outcome using clinical variables. Additional models are built that also include cluster labels (all or individual cluster labels), and cross-validation performance is measured. Outcomes are grouped by the metric used, and the dose threshold considered (3,5,7). This view was removed as our collaborators preferred to rely on traditional statistics with fewer thresholds when deciding on outcomes.

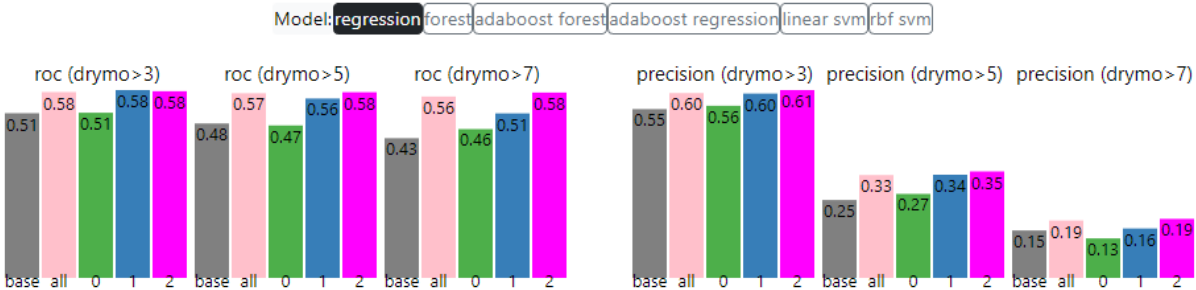

Fig B5. Early version of the interface using patient-specific view and star-charts to encode symptom ratings for patients and clusters. The patient view (top-right) used a single selected patient (133) within a selected cluster at the top. The additional patients on the left are the most similar within the same cluster, while the patients on the left show similar patients in different clusters, based on treatment categories and doses. Patients in the cluster are encoded with a user-selected dose value (e.g. V55) to each organ in red, while patients in other clusters are color-encoded with the difference in dose values from the selected patient using a blue-green color scale, where green indicates higher dose, and blue indicates lower doses. Star charts encode patient symptoms. This view was removed as we found that inspecting patients via a tooltip on the scatterplot was more efficient.

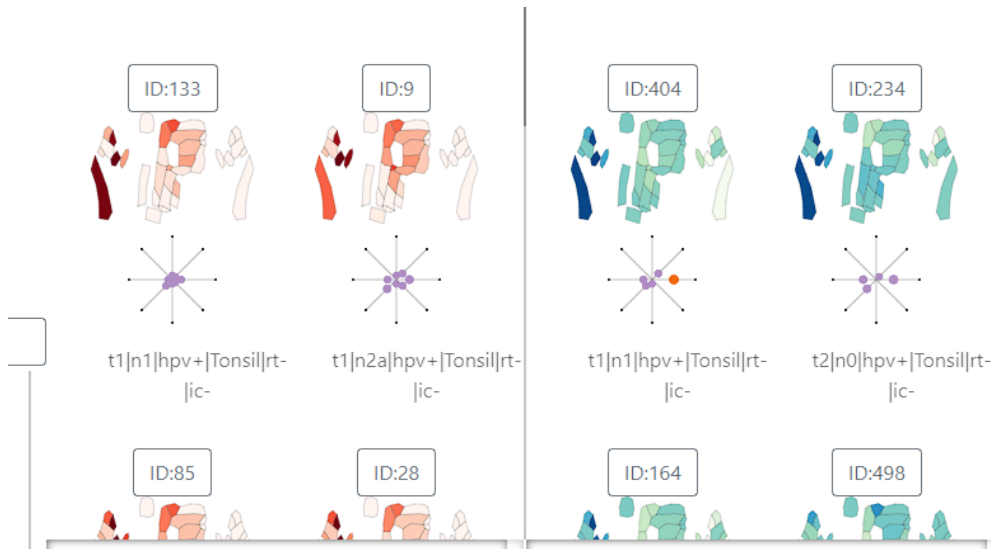

Fig B6. Alternative Variants of the patient encodings that use a color gradient to show the full dose-histogram within each organ. The outer layer represents the V5 while the innermost layer represents the V75 to the respective organ. Symptoms at different stages (baseline, during treatment, and post-treatment) are shown as bar charts using small multiples for a set of symptoms of interest. Patient color encodings for the counterfactual patients show absolute rather than relative doses.

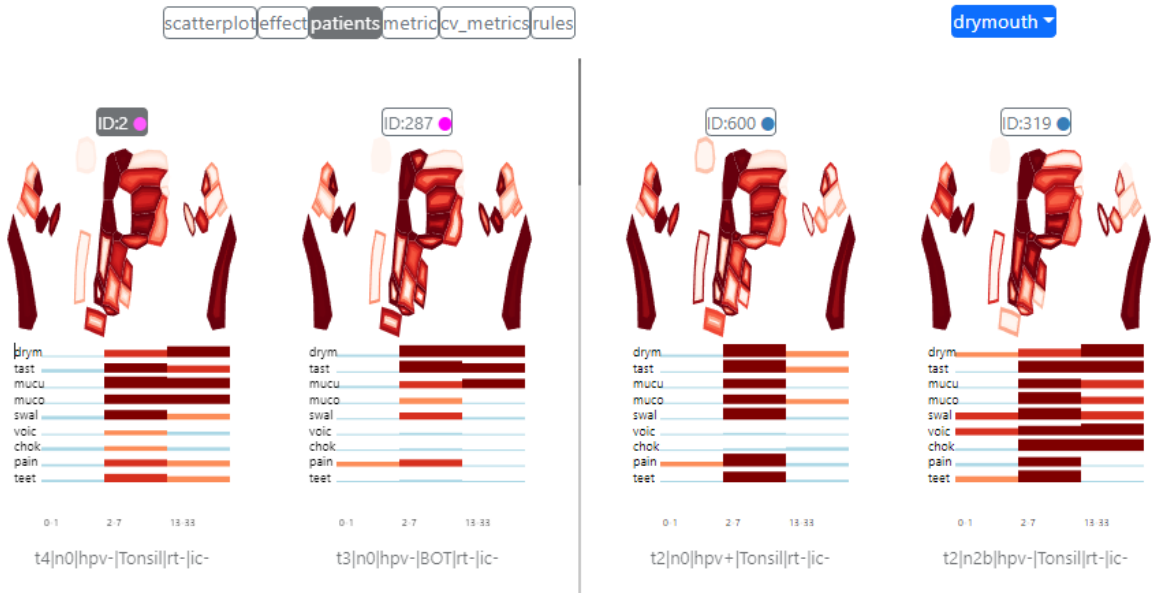

Fig B7. Alternative Version of the scatterplot that used shape instead of glyphs to encode symptom ratings. Circles, triangles, and diamonds encoded patients with symptom ratings of < 5, 5-7, or >7, respectively. Shape-encodings were changed to glyphs as our collaborators felt that the encodings were unintuitive.

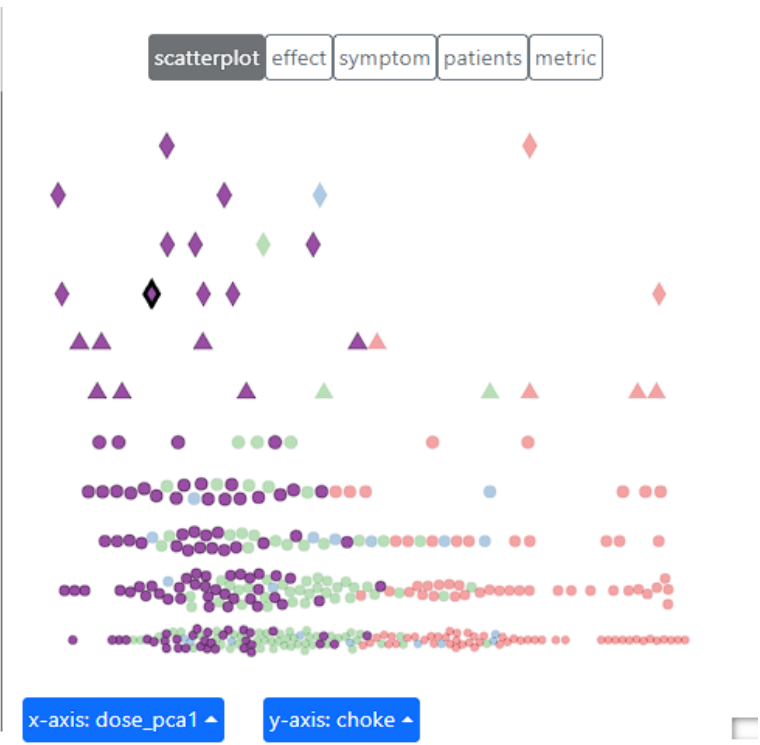

Fig B8. early version of the cluster view. The view on the right encodes cluster doses as in the final version. A star chart on the center-right encodes symptom ratings within the cluster, where the dot encodes median rating and the bold line indicates the 25-75% confidence interval. Circle size encodes odds-ratio for the cluster to experience severe symptoms. Finally, histograms on the right encode the distribution of clinical features within each cluster. The symptom charts were replaced with our current symptom plots as we wanted to focus on comparing trajectories between clusters. Clinical feature plots were removed to save space, as we did not use the information in the development of clusters outside of using them as covariates in our predictive models. Instead, extensive demographic distributions were calculated after clusters were finalized and thus did not need to be included in the interactive system.

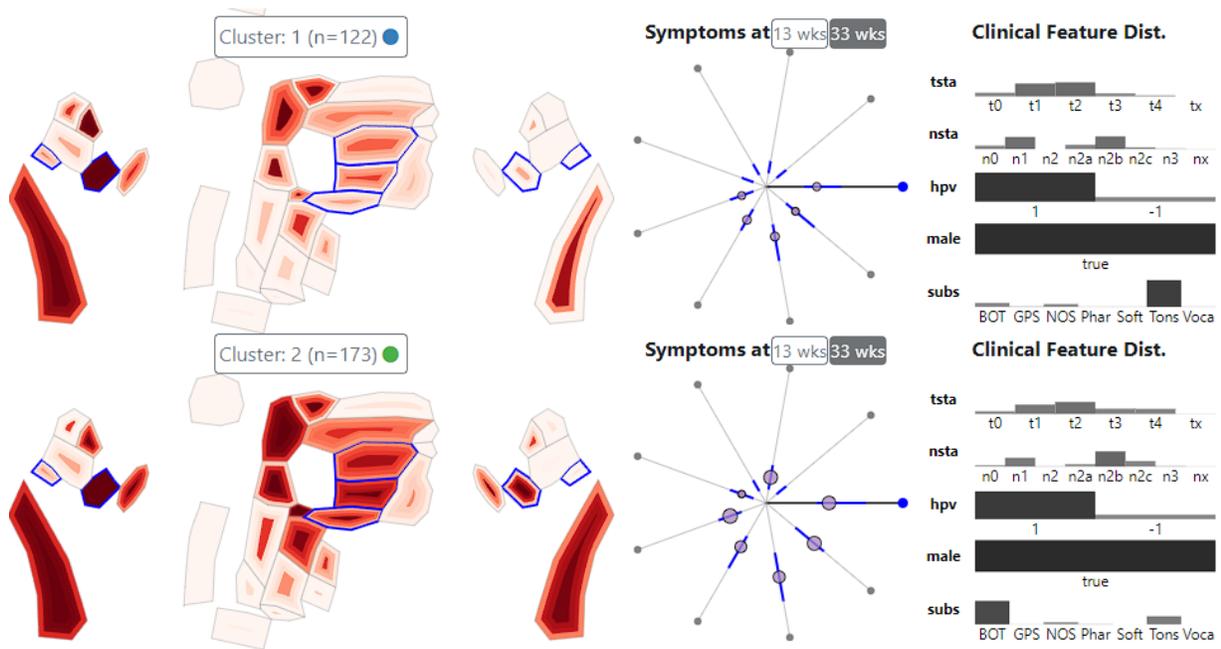

Fig B9. Early design of the layout of DASS in different configurations. The original layout was divided into 5 views: a control panel (top-left), the cluster view (center left), the symptom trajectory view (bottom-left), and two windows on the left that allow the user to toggle between views. Several adjustments to the layout and color schemes were made to de-clutter the interface based on feedback from collaborators. We also incorporated control-panels for each view into titles for each plot in order to make them more readable.

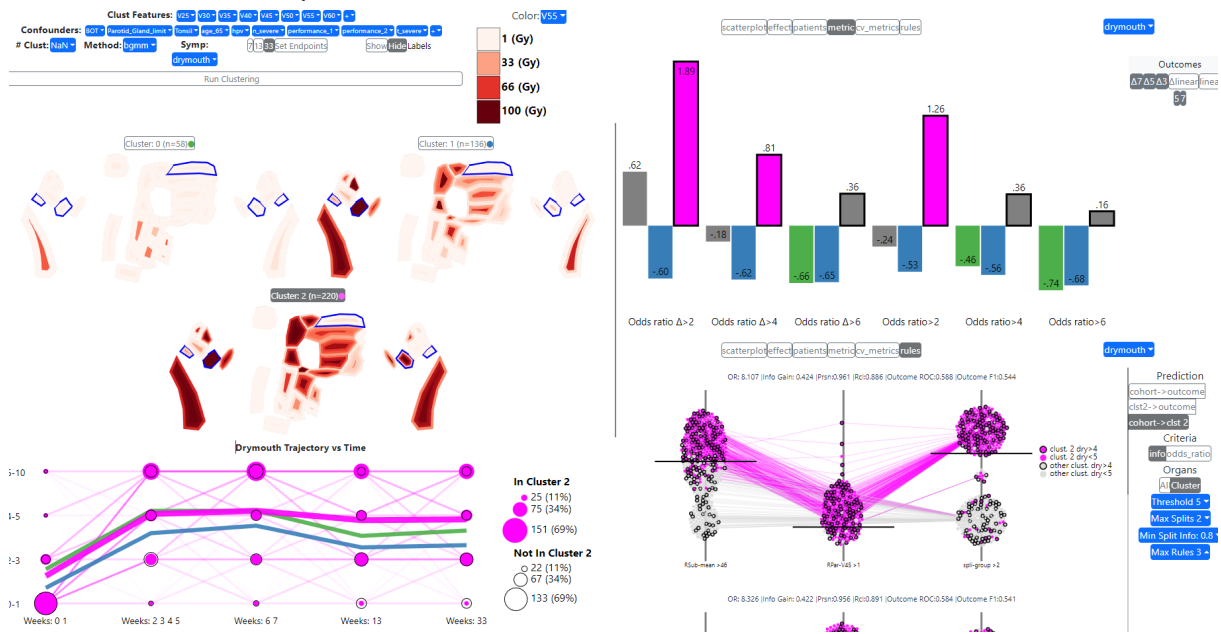

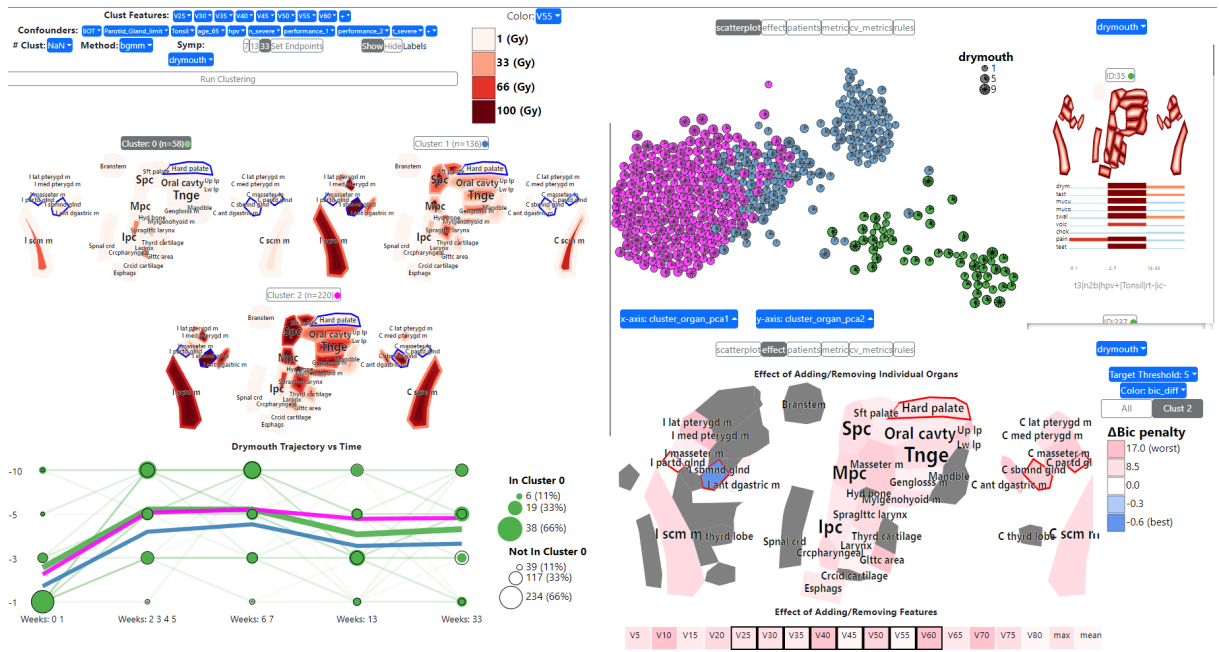

## C. Cluster Quantitative Analysis

We include here a comparison of the performance of the clusters contained using our system to baseline clusters that are obtained via methods without visual steering. In this experiment, we use a baseline method of generating clusters to predict patient outcomes without use of the DASS interface. We then compare those results to clusters made from the Dass interface using two kinds of starting parameters: a default set determined by treatment guidelines, and the parameters used in the baseline clusters. Clusters are evaluated using a simple ablation study where a regression model is built to predict the desired outcome with and without the cluster label to test the effect of the model on cross validation model performance. We perform this experiment on three outcomes: late severe (rating >5) drymouth, swallow dysfunction, and choking.

To generate baseline clusters, we tested the results of clusters generated through a combination of input features and organs with 2-4 clusters, using a bayesian mixture model. Because we require that the organs in the model be linked to the outcome, we compiled lists of potential organ combinations for each outcome based on clinical literature. For drymouth, we consider the set of organs taken from NTCP models built for predicting xerostomia [1], and for swallow and choke, we use organs taken from NTCP models built for dysphagia [2]. In addition, we included sets of organs gathered from clinician feedback on the organs they believe would be related to each.

To compare our models, we used two variants of clusters that are obtained from DASS for each outcome. As a baseline, we found clusters following the regular approach, where we used a general starting point of the most important organ of interest. In addition, we consider models

that are obtained from Dass by using the model parameters in the baseline models found using grid-search. Simplified clusters for all cases are generated as described in the main text. The parameters of the final clusters are listed in Table C1.

For quantitative analysis, we convert our clusters into a clinical stratification during cross-validation as follows: first, we rank each cluster based on the number of patients that experience the given outcome in the training data split and assign risk to patients in the test data based on the rank of their clusters, where the highest-risk cluster is given a risk score of 1, the second highest-risk cluster is given a score of (number of clusters - 2)/(number of clusters - 1), etc, and the lowest risk is given a score of 0. For the simplified cluster, we always assign a risk of 1 to the high-dose cluster and 0 otherwise. We then calculate the cross-validation mean Area-under the curve score (AUC), and the Mathews correlation coefficient (MCC) for predicting each outcome for each cluster model. In addition, we calculate the results from using only the highest-dose (HD) cluster for each stratification, as this is a likely use case for clinical applications when deciding whether to prescribe preventative treatment. Thus, we generate AUC and MCC scores for 3 different cases for each clustering.

Results for AUC are shown in Figure C1. Drymouth clusters obtained using DASS, the main symptom of interest that we used to build the model, perform better for all measures compared to grid-search. Choke clusters perform better in general, although interestingly, grid-search simplified explanations tend to outperform other rule-based simplified explanations. For swallow, the second Dass clusters outperform the gridsearch cluster for the stratification but not the rule-based clusters.

We note that while our swallow and choke clusters have mixed results, we can reasonably beat the baseline clusters for several measures without the need for extensive literature review to identify the optimal organs to consider when testing multiple clusters, in addition to the general benefits of interactive changing of the clusters to balance between other considerations such as the maximum number of clusters or comparison of performance across multiple thresholds. Drymouth results tended to be the most consistent, most likely due to the higher rates of severe drymouth, while severe swallow and choke are relatively rare, making the cluster predictions difficult. In addition, the organs linked to drymouth tend to be salivary glands rather than muscles, and are likely more sensitive to lower doses.

Table C1. Table of features used in each cluster for our analysis. The prefixes 'Lt' and "Rt" refer to the ipsilateral (same side as main tumor) and contralateral (opposite side as main tumor) organ, respectively. Legend: Dass: Models built using all clusters using DASS with a default starting point; Alt Dass: Models built using DASS with the baseline clusters at a starting point. Baseline: Models built using clusters obtained through standard grid search.

| Symptom | organs | N Clusters | Organs | Dose Features | Simple Rule |
|---|---|---|---|---|---|
| Drymouth | Dass | 3 | Rt_Parotid_Gland, Lt_Parotid_Gland, Rt_Submandibular_Gland, Lt_Submandibular_Gland, Hard_Palate | V25-V60 | V55_Rt_Submandibular_Gland>23.0, V65_Lt_Submandibular_Gland>51.0, V10_Rt_Parotid_Gland>18.0 |
| | Dass Alt | 3 | Rt_Parotid_Gland, Lt_Parotid_Gland, Rt_Submandibular_Gland, Lt_Submandibular_Gland, Soft_Palate | V30-V55 | V55_Rt_Submandibular_Gland>23.0, V65_Lt_Submandibular_Gland>51.0, V10_Rt_Parotid_Gland>18.0 |
| | Baseline | 3 | Lt_Parotid_Gland, Rt_Parotid_Gland, Lt_Submandibular_Gland, Rt_Submandibular_Gland, Soft_Palate, Upper_Lip, Lower_Lip, Extended_Oral_Cavity, Mylogeniohyoid_M | mean dose, max dose | V60_Mylogeniohyoid_M>32.0, V55_Extended_Oral_Cavity>28.0, V30_Rt_Submandibular_Gland>62.0, V65_Lt_Submandibular_Gland>1.0 |
| Choke | Dass | 3 | IPC, MPC, Supraglottic_Larynx, Esophagus, Mylogeniohyoid_M | V30-V65 | V60_Supraglottic_Larynx>40.0, V55_MPC>17.0, V55_Mylogeniohyoid_M>24.0, V15_Esophagus>22.0 |
| | Dass Alt | 3 | IPC, MPC, Supraglottic_Larynx, Cricopharyngeal_Muscle, Mylogeniohyoid_M | V30, V40, V50, V60 | mean_dose_IPC>49.0, mean_dose_MPC>54.0, mean_dose_Cricopharyngeal_Muscle>16.0 |
| | Baseline | 3 | SPC, IPC, Supraglottic_Larynx, Rt_Parotid_Gland, Cricopharyngeal_Muscle | mean dose, max dose | V5_Rt_Parotid_Gland>72.0, V65_SPC>3.0, V55_Supraglottic_Larynx>46.0, max_dose_IPC>46.0 |
| Swallow | Dass | 3 | IPC, MPC, SPC' Supraglottic_Larynx, Esophagus | V35-V70 | V60_Supraglottic_Larynx>40.0, V65_SPC>3.0, V15_Esophagus>22.0, |

| | | | | | V55_MPC>17.0 |
|---|---|---|---|---|---|
| | Dass Alt | 2 | IPC, SPC, Cricopharyngeal_Muscle, Rt_Parotid_Gland | V65, max dose | max_dose_Cricopharyngeal_Muscle>63.0 |
| | Baseline | 2 | SPC, IPC, Supraglottic_Larynx, Rt_Parotid_Gland, Cricopharyngeal_Muscle | mean_dose, V40-V65 | V5_Rt_Parotid_Gland>72.0, V65_SPC>3.0, V55_Supraglottic_Larynx>46.0, max_dose_IPC>46.0 |

Figure C1. AUC and MCC scores for all model types. For cluster stratifications, Alt Dass clusters and high-dose clusters outperform the baseline models. Alt Dass performs better than Dass for Choke and Swallow, but not Drymouth.

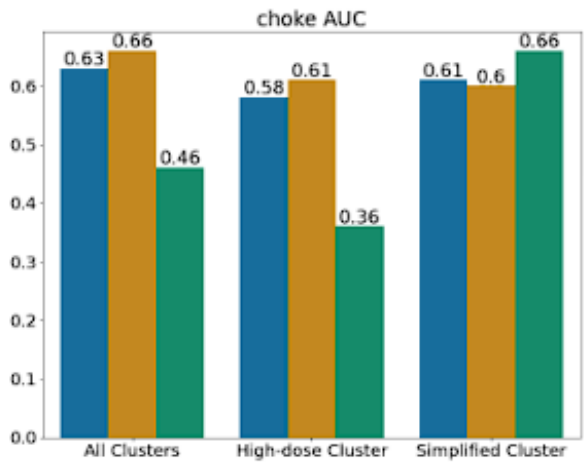
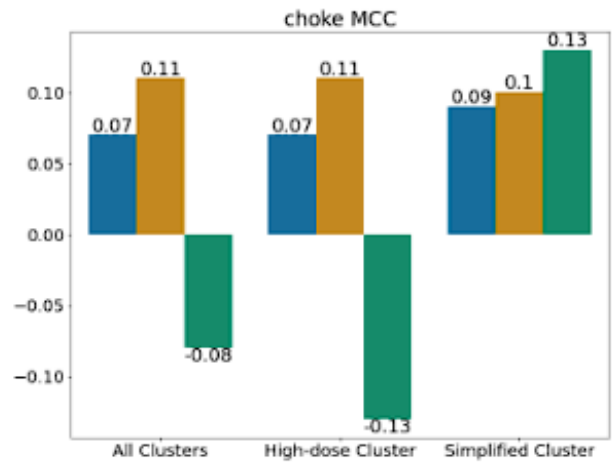
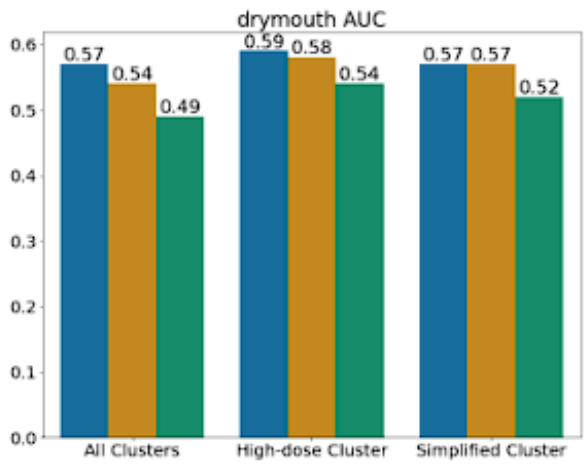
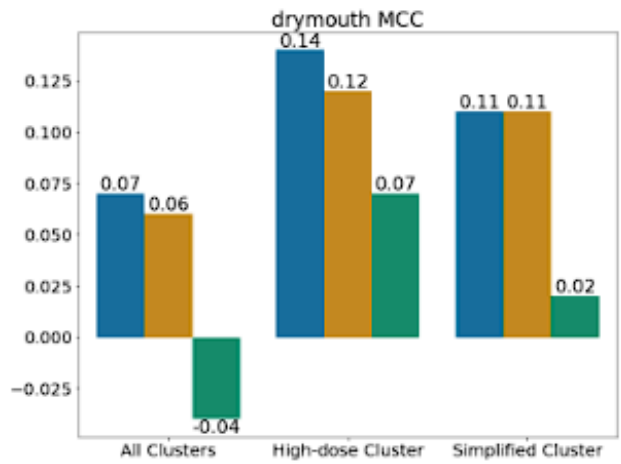
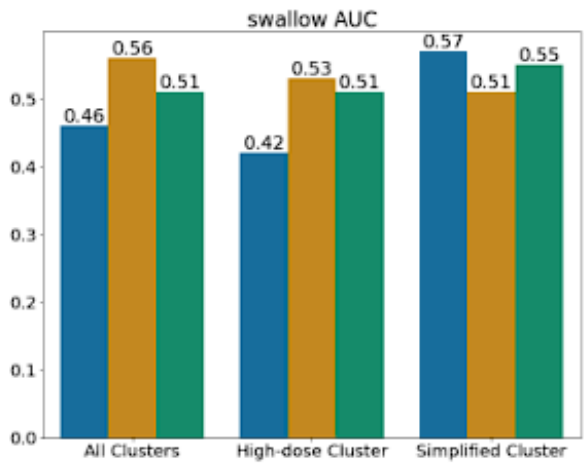
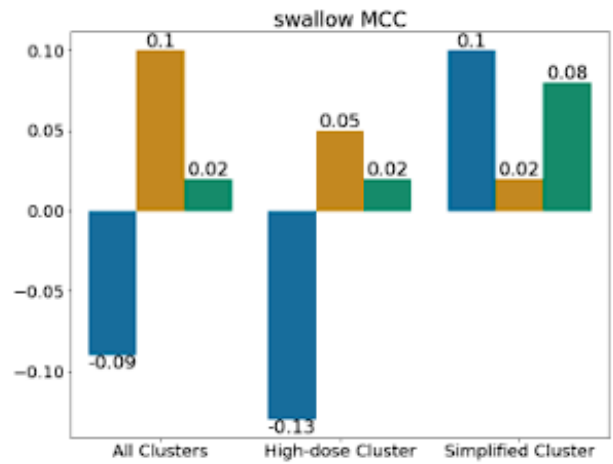

Table C2. Results from 5-fold cross-validation of different stratification models for predicting severe late symptoms. We consider both absolute ratings > 4 at 6 months, and increase in ratings > 4 from baseline at 6 months. Models include stratification with all clusters (stratification), stratification using only the highest dose clusters (High-dose), and the simplified high-risk rule explanations. Legend: Dass: Models built using all clusters using DASS with a default starting point; Alt Dass: Models built using DASS with the baseline clusters at a starting point. Baseline: Models built using clusters obtained through standard grid search.

| Symptom | Change From Baseline | Model | ROC-AUC | MCC | Precision | Recall | F1 |
|---|---|---|---|---|---|---|---|
| | | Dass Stratification | 0.628 | 0.070 | 0.059 | 0.500 | 0.106 |
| | | Alt Dass Stratification | 0.659 | 0.106 | 0.080 | 0.429 | 0.135 |
| | | Baseline Stratification | 0.462 | -0.082 | 0.026 | 0.357 | 0.048 |
| | | Dass High-dose | 0.584 | 0.070 | 0.059 | 0.500 | 0.106 |
| | | Alt Dass High-dose | 0.611 | 0.106 | 0.080 | 0.429 | 0.135 |
| | | Baseline High-dose | 0.357 | -0.126 | 0.000 | 0.000 | 0.000 |
| | | Dass Simplified | 0.611 | 0.088 | 0.060 | 0.643 | 0.110 |
| | | Al Dass Simplified | 0.596 | 0.100 | 0.083 | 0.357 | 0.135 |
| | FALSE | Baseline Simplified | 0.662 | 0.127 | 0.066 | 0.786 | 0.122 |
| | | Dass Stratification | 0.652 | 0.065 | 0.051 | 0.500 | 0.092 |
| | | Alt Dass Stratification | 0.714 | 0.131 | 0.080 | 0.500 | 0.138 |
| | | Baseline Stratification | 0.469 | -0.085 | 0.021 | 0.333 | 0.039 |
| Choke | TRUE | | | | | | |

|   |   |   | Condition | V1 | V2 | V3 | V4 | V5 |
|---|---|---|---|---|---|---|---|---|
| | | | Dass High-dose | 0.584 | 0.065 | 0.051 | 0.500 | 0.092 |
| | | | Alt Dass High-dose | 0.648 | 0.131 | 0.080 | 0.500 | 0.138 |
| | | | Baseline High-dose | 0.358 | -0.116 | 0.000 | 0.000 | 0.000 |
| | | | Dass Simplified | 0.623 | 0.090 | 0.053 | 0.667 | 0.099 |
| | | | Al Dass Simplified | 0.627 | 0.122 | 0.083 | 0.417 | 0.139 |
| | | | Baseline Simplified | 0.685 | 0.135 | 0.060 | 0.833 | 0.112 |
| | | | Dass Stratification | 0.571 | 0.065 | 0.220 | 0.574 | 0.318 |
| | | | Alt Dass Stratification | 0.540 | 0.058 | 0.216 | 0.603 | 0.318 |
| | | | Baseline Stratification | 0.490 | -0.037 | 0.179 | 0.426 | 0.252 |
| | | | Dass High-dose | 0.586 | 0.137 | 0.244 | 0.691 | 0.360 |
| | | | Alt Dass High-dose | 0.577 | 0.124 | 0.236 | 0.706 | 0.354 |
| | | | Baseline High-dose | 0.542 | 0.069 | 0.215 | 0.706 | 0.330 |
| | | | Dass Simplified | 0.568 | 0.108 | 0.241 | 0.574 | 0.339 |
| | | | Al Dass Simplified | 0.568 | 0.108 | 0.241 | 0.574 | 0.339 |
| | FALSE | | Baseline Simplified | 0.515 | 0.025 | 0.208 | 0.382 | 0.269 |
| | | | Dass Stratification | 0.583 | 0.046 | 0.141 | 0.568 | 0.226 |
| | | | Alt Dass Stratification | 0.546 | 0.035 | 0.137 | 0.591 | 0.222 |
| | | | Baseline Stratification | 0.485 | -0.059 | 0.105 | 0.386 | 0.165 |
| | | | Dass High-dose | 0.613 | 0.150 | 0.171 | 0.750 | 0.278 |
| | | | Alt Dass High-dose | 0.596 | 0.130 | 0.163 | 0.750 | 0.267 |
| | | | Baseline High-dose | 0.551 | 0.070 | 0.143 | 0.727 | 0.240 |
| | | | Dass Simplified | 0.573 | 0.097 | 0.160 | 0.591 | 0.252 |
| Drymouth | TRUE | | | | | | | |

| | | | | | | | |
|---|---|---|---|---|---|---|---|
| | | Al Dass Simplified | 0.573 | 0.097 | 0.160 | 0.591 | 0.252 |
| | | Baseline Simplified | 0.516 | 0.022 | 0.136 | 0.386 | 0.201 |
| | FALSE | Dass Stratification | 0.465 | -0.091 | 0.140 | 0.393 | 0.206 |
| | | Alt Dass Stratification | 0.560 | 0.104 | 0.247 | 0.328 | 0.282 |
| | | Baseline Stratification | 0.513 | 0.024 | 0.180 | 0.770 | 0.292 |
| | | Dass High-dose | 0.418 | -0.125 | 0.122 | 0.311 | 0.175 |
| | | Alt Dass High-dose | 0.533 | 0.053 | 0.190 | 0.689 | 0.298 |
| | | Baseline High-dose | 0.512 | 0.021 | 0.179 | 0.770 | 0.291 |
| | | Dass Simplified | 0.567 | 0.102 | 0.218 | 0.557 | 0.313 |
| | | Al Dass Simplified | 0.511 | 0.018 | 0.179 | 0.754 | 0.289 |
| | | Baseline Simplified | 0.550 | 0.075 | 0.205 | 0.557 | 0.300 |
| | | Dass Stratification | 0.506 | -0.021 | 0.116 | 0.465 | 0.186 |
| | | Alt Dass Stratification | 0.538 | 0.083 | 0.173 | 0.326 | 0.226 |
| | | Baseline Stratification | 0.523 | 0.040 | 0.130 | 0.837 | 0.225 |
| | | Dass High-dose | 0.431 | -0.092 | 0.090 | 0.326 | 0.141 |
| | | Alt Dass High-dose | 0.510 | 0.014 | 0.127 | 0.651 | 0.212 |
| | | Baseline High-dose | 0.494 | -0.011 | 0.122 | 0.837 | 0.212 |
| | | Dass Simplified | 0.577 | 0.101 | 0.160 | 0.581 | 0.251 |
| | | Al Dass Simplified | 0.504 | 0.007 | 0.125 | 0.744 | 0.213 |
| Swallow | TRUE | Baseline Simplified | 0.560 | 0.079 | 0.151 | 0.581 | 0.239 |

Table C3. Distribution of outcomes in the cohort considered.

| Symptom | Total >4 at 6M | Mean Rating | 95% CI |
|---|---|---|---|
| Drymouth | 153 (43.84%) | 4.338109 | 0.4-9 |
| Swallow | 46 (12.18%) | 2.137536 | 0-7 |
| Choke | 19 (5.44%) | 1.111748 | 0-5 |

Table C4. Parameters used in the gridsearch for the baseline. All parameters were tested with a gaussian mixture models using 2-4 components

| Dose Values (All Outcomes) |
|---|
| ['mean_dose','V25','V30','V35','V40','V45','V50','V55','V60','V65','V70','V75','V80'],<br>['mean_dose','V25','V30','V35','V40','V45','V50'],<br>['mean_dose','V25','V30','V35','V40'],<br>['mean_dose','V30','V35','V40','V45','V50','V55','V60','V65','V70','V75','V80'],<br>['mean_dose','V40','V45','V50','V55','V60','V65','V70','V75','V80'],<br>['mean_dose','V55','V60','V65','V70','V75','V80'],<br>['mean_dose','V60','V65','V70','V75','V80'],<br>['mean_dose','V40','V45','V50','V55','V60','V65'],<br>['mean_dose','V40','V45','V50','V55','V60'],<br>['V25','V30','V35','V40','V45','V50','V55','V60','V65','V70','V75','V80'],<br>['V25','V30','V35','V40','V45','V50'],<br>['V25','V30','V35','V40'],<br>['V30','V35','V40','V45','V50','V55','V60','V65','V70','V75','V80'],<br>['V40','V45','V50','V55','V60','V65','V70','V75','V80'],<br>['V55','V60','V65','V70','V75','V80'],<br>['V60','V65','V70','V75','V80'],<br>['V40','V45','V50','V55','V60','V65'],<br>['V40','V45','V50','V55','V60'],<br>['mean_dose'],<br>['max_dose'],<br>['max_dose','mean_dose'] |
| Organs (Drymouth) |

| |
|---|
| ['Tongue', 'Mylogeniohyoid_M', 'Genioglossus_M','Rt_Parotid_Gland','Lt_Parotid_Gland', 'Rt_Submandibular_Gland','Lt_Submandibular_Gland', 'Soft_Palate', 'Extended_Oral_Cavity','Supraglottic_Larynx','Larynx'] |
| ['Lt_Parotid_Gland','Rt_Parotid_Gland','Lt_Submandibular_Gland','Rt_Submandibular_Gland','Soft_Palate', 'Upper_Lip','Lower_Lip', 'Extended_Oral_Cavity','Mylogeniohyoid_M'] |
| Organs (Swallow) |
| ['SPC','IPC','Supraglottic_Larynx','Rt_Parotid_Gland','Cricopharyngeal_Muscle'] |
| ['IPC','MPC', 'SPC', 'Mylogeniohyoid_M','Tongue'] |
| ['SPC','IPC','MPC','Supraglottic_Larynx', 'Rt_Parotid_Gland','Cricopharyngeal_Muscle'] |
| Organs (Choke) |
| ['SPC','IPC','Supraglottic_Larynx','Rt_Parotid_Gland','Cricopharyngeal_Muscle'] |
| ['Rt_Masseter_M', 'Lt_Masseter_M', 'Rt_Medial_Pterygoid_M', 'Lt_Medial_Pterygoid_M', 'Rt_Lateral_Pterygoid_M', 'Lt_Lateral_Pterygoid_M'] |
| ['Rt_Masseter_M','Lt_Masseter_M','Rt_Medial_Pterygoid_M', 'Lt_Medial_Pterygoid_M', 'Rt_Lateral_Pterygoid_M', 'Lt_Lateral_Pterygoid_M', 'Supraglottic_Larynx', 'Larynx','Glottic_Area','Thyroid_cartilage', 'Cricopharyngeal_Muscle', 'Cricoid_cartilage','Esophagus' ] |

## D. Qualitative Analysis Results

Table D1. Results From the usability evaluation questionnaire. Ratings are given on a range of 1 (not useful) to 5 (very useful).

| | Additive Effects | Dose Clusters | Scatterplot | Outcome Plots | Rule Mining | Overall |
|---|---|---|---|---|---|---|

|  | Useful | Usable | Usefule | Usable | Usefule | Usable | Usefule | Usable | Explanations Usefule | Usable | Usefule | Usable |
|---|---|---|---|---|---|---|---|---|---|---|---|---|
| P1 | 4 | 4 | 5 | 4 | 5 | 4 | 5 | 4 | 5 | 4 | 5 | 4 |
| P2 | 5 | 5 | 5 | 5 | 5 | 5 | 4 | 5 | 5 | 4 | 5 | 5 |
| P3 | 5 | 5 | 5 | 5 | 5 | 5 | 4 | 3 | 4 | 4 | 5 | 5 |
| P4 | 5 | 5 | 5 | 5 | 5 | 5 | 4 | 4 | 4 | 4 | 5 | 5 |
| P5 | 5 | 4 | 5 | 3 | 5 | 5 | 5 | 5 | 5 | 4 | 4 | 4 |
| P6 | 5 | 4 | 3 | 3 | 4 | 4 | 5 | 4 | 5 | 5 | 4 | 4 |
| P7 | 4 | 3 | 2 | 2 | 4 | 4 | 4 | 1 | 5 | 4 | 4 | 2 |
| P8 | 5 | 5 | 5 | 4 | 4 | 4 | 5 | 5 | 5 | 5 | 5 | 5 |
| Average | 4.75 | 4.38 | 4.38 | 3.88 | 4.63 | 4.50 | 4.50 | 3.88 | 4.75 | 4.25 | 4.63 | 4.25 |

Figure D1. Plot of user responses to the usability questionnaire.

Open ended comments

**P6:** The Dose Cluster Panel a bit complex at the beginning. But became clearer later on. Use cases are well explained in the end.

For the additive effects panel, it's easy to see the effect of adding other organs but I believe there is a factor of domain knowledge missing. For instance, adding an organ may look like a good idea but it may not be relevant to the symptom being investigated (based on AW's comments during presentation).

**P7:** The Dose Cluster panel encoding confused me because the distribution was mapped to spatial locations, and that matching was really trying to encode quantiles. Also it's quantized but the line chart is not.

I see a lot of potential in making the interface more actionable than descriptive. For instance, what's the effect of removing outliers or so, and comparing the cluster afterwards. Not sure if that's the purpose though

I'm not very convinced by the choice of the force layout graphs because they mutate the positions of the scatterplots or swarm for the sake of visibility. In doing so, they add some uncertainty of whether the location is accurate or not.

I liked that the outcome plots show what results are significant, but the p value could be parametrized given the different opinions of what threshold is actually significant. For the temporal trajectory, the choose of size and whether a circle is bigger than another is difficult for me to parse. I correlate the radius size with bigger being better or worse, but I believe it encodes population, and the relationship between which circle includes another circle is actually what matters. I cannot still read that well.

I liked the rule mining pcp the most, but didn't like the force layout for reasons described above. I think it could have more screen space to be able to add more rules if needed.



\setlength{\voffset}{-2.54cm}
\setlength{\hoffset}{-2.54cm}
\end{document}